\documentclass[sigconf]{acmart}
\acmConference[Anonymous Submission to ACM CCS 2024]{ACM Conference on Computer and Communications Security}{14-19 November 2024}{Copenhagen, Denmark}
\def\BibTeX{{\rm B\kern-.05em{\sc i\kern-.025em b}\kern-.08emT\kern-.1667em\lower.7ex\hbox{E}\kern-.125emX}}

\def\B{BlueSWAT}

\def\M{manufacturers}

\usepackage{booktabs}
\usepackage{lscape}
\usepackage{
  color,
  float,
  epsfig,
  wrapfig,
  graphics,
  graphicx,
  subcaption,
}
\usepackage{textcomp}
\usepackage{courier}
\usepackage{setspace}
\usepackage{latexsym,fancyhdr,url}
\usepackage{enumerate}
\usepackage{algorithm2e}
\usepackage{algpseudocode}
\usepackage{graphics}
\usepackage{xparse} 
\usepackage{xspace}
\usepackage{multirow}
\usepackage{csvsimple}
\usepackage{balance}

\usepackage{placeins}
\usepackage{soul,color}
\sethlcolor{cyan}

\usepackage{
  tikz,
  pgfplots,
  pgfplotstable
}
\usepackage{hyperref}

\usetikzlibrary{
  shapes.geometric,
  arrows,
  external,
  pgfplots.groupplots,
  matrix
}

\usepackage{listings} 
\usepackage{pifont}

\pgfplotsset{compat=1.9}

\usepackage{mathtools,}

\DeclareMathAlphabet{\mathcal}{OMS}{cmsy}{m}{n}

\DeclareGraphicsExtensions{%
    .png,.PNG,%
    .pdf,.PDF,%
    .jpg,.mps,.jpeg,.jbig2,.jb2,.JPG,.JPEG,.JBIG2,.JB2}

\usepackage{xparse}
\newcommand{\bnm}{\begin{newmath}}
\newcommand{\enm}{\end{newmath}}

\newcommand{\bea}{\begin{eqnarray*}}%
\newcommand{\eea}{\end{eqnarray*}}%

\newcommand{\bne}{\begin{newequation}}
\newcommand{\ene}{\end{newequation}}

\newcommand{\bal}{\begin{newalign}}
\newcommand{\eal}{\end{newalign}}

\newenvironment{newalign}{\begin{align}%
\setlength{\abovedisplayskip}{4pt}%
\setlength{\belowdisplayskip}{4pt}%
\setlength{\abovedisplayshortskip}{6pt}%
\setlength{\belowdisplayshortskip}{6pt} }{\end{align}}

\newenvironment{newmath}{\begin{displaymath}%
\setlength{\abovedisplayskip}{4pt}%
\setlength{\belowdisplayskip}{4pt}%
\setlength{\abovedisplayshortskip}{6pt}%
\setlength{\belowdisplayshortskip}{6pt} }{\end{displaymath}}

\newenvironment{newequation}{\begin{equation}%
\setlength{\abovedisplayskip}{4pt}%
\setlength{\belowdisplayskip}{4pt}%
\setlength{\abovedisplayshortskip}{6pt}%
\setlength{\belowdisplayshortskip}{6pt} }{\end{equation}}

\newcounter{ctr}

\newcounter{mytable}
\def\mytable{\begin{centering}\refstepcounter{mytable}}
\def\endmytable{\end{centering}}

\newcounter{myfig}
\def\myfig{\begin{centering}\refstepcounter{myfig}}
\def\endmyfig{\end{centering}}

\newlength{\saveparindent}
\newlength{\saveparskip}
\newcommand{\E}{{\rm I\kern-.3em E}}

\renewcommand{\eqref}[1]{\mbox{Equation~(\ref{#1})}}

\def \part {part}

\renewcommand{\paragraph}[1]{\vspace*{6pt}\noindent\textbf{#1}\;}

\def \blackslug{\hbox{\hskip 1pt \vrule width 4pt height 8pt
    depth 1.5pt \hskip 1pt}}
\def \qed{\quad\blackslug\lower 8.5pt\null\par}

\newcounter{mynote}[section]

\newcommand\ignore[1]{}

\newcounter{rcnote}[section]

\newcounter{mrnote}[section]

\newcounter{fknote}[section]

\newcounter{anote}[section]

\DeclareMathSymbol{\mlq}{\mathord}{operators}{``}
\DeclareMathSymbol{\mrq}{\mathord}{operators}{`'}

\newcommand{\rhf}[2]{R_{f, \gamma}}

\DeclareDocumentCommand{\edist}{o o}{
  \ensuremath{
    \IfNoValueTF{#1}{{d}}{{\sf d}(#1,#2)}
  }
}

\newcommand{\olrk}[1]{\ifx\nursymbol#1\else\!\!\mskip4.5mu plus 0.5mu\left(\mskip0.5mu plus0.5mu #1\mskip1.5mu plus0.5mu \right)\fi}

\NewDocumentCommand{\indseq}{ O{1} O{r} }{{#1}\ldots {#2}}

\usepackage{makecell}
\usepackage{fontawesome}
\usepackage{bbding}
\usepackage{wasysym}

\begin{document}

\fancyhead{}
\def\thetitle{\B : A Lightweight State-Aware Security Framework for Bluetooth Low Energy}
\title{\thetitle}

\author{Xijia Che}
\affiliation{Tsinghua University}
\affiliation{cxj22@mails.tsinghua.edu.cn}

\author{Yi He}
\affiliation{Tsinghua University}
\affiliation{heyi21@mails.tsinghua.edu.cn}

\author{Xuewei Feng}
\affiliation{Tsinghua University}
\affiliation{fengxw06@126.com}

\author{Kun Sun}
\affiliation{George Mason University}
\affiliation{ksun3@gmu.edu}

\author{Ke Xu}
\affiliation{Tsinghua University}
\affiliation{xuke@tsinghua.edu.cn}

\author{Qi Li}
\affiliation{Tsinghua University}
\affiliation{qli01@tsinghua.edu.cn}

\date{}

\begin{abstract}

Bluetooth Low Energy (BLE) is a short-range wireless communication technology for resource-constrained IoT devices. Unfortunately, BLE is vulnerable to session-based attacks, where previous packets construct exploitable conditions for subsequent packets to compromise connections. Defending against session-based attacks is challenging because each step in the attack sequence is legitimate when inspected individually. In this paper, we present \B , a lightweight state-aware security framework for protecting BLE devices. To perform inspection on the session level rather than individual packets, \B\ leverages a finite state machine (FSM) to monitor sequential actions of connections at runtime. Patterns of session-based attacks are modeled as malicious transition paths in the FSM. To overcome the heterogeneous IoT environment, we develop a lightweight eBPF framework to facilitate universal patch distribution across different BLE architectures and stacks, without requiring device reboot. We implement \B\ on 5 real-world devices with different chips and stacks to demonstrate its cross-device adaptability. On our dataset with 101 real-world BLE vulnerabilities, \B\ can mitigate 76.1\% of session-based attacks, outperforming other defense frameworks. In our end-to-end application evaluation,  \B\ patches introduce an average of 0.073\% memory overhead and negligible latency.



\end{abstract}
\maketitle
\keywords{LaTeX template, ACM CCS, ACM}

\section{Introduction}
\label{sec:intro}

Bluetooth is a widely used wireless communication protocol in the Internet of Things (IoT), allowing smart devices to exchange data in short range.
In 2010, the Bluetooth Core Specification 4.0~\cite{btspec4} heralds the advent of BLE, an innovative wireless protocol designed specifically for resource-limited IoT devices. BLE, which focuses on low-power communication, has experienced substantial growth over the past decade. More than 5 million BLE-enabled smart devices are estimated to be in use by 2023, and total annual shipments of BLE devices will reach 7 billion by 2027~\cite{2023BluetoothMarketUpdate}.

With the proliferation of BLE-enabled devices in smart homes and wearables, a vast amount of sensitive information is being transmitted over BLE channels and the attacks continue to emerge, including the \textit{packet-based attacks} and the \textit{session-based attacks}.   
A packet-based attack \cite{Garbelini2020SweynToothUM, cve202010061,10066,10069,3433,10065} manipulates one crafted packet to exploit the target BLE stack, which usually aims at implementation flaws such as missing bounds checks. It may lead to device DoS or even remote code execution (RCE) when carefully manipulated.
By contrast, a session-based attack \cite{Claverie2021BlueMirrorRO,Wu2022FormalMD,Wu2020BLESASA,Zhang2020BreakingSP,Nguyen2014FormalAO,Tschirschnitz2021MethodCA,bleedingbit,Xie2023AccessYT} is performed by constructing complex BLE interaction processes that require multiple previous data packets to lay the groundwork for the final attack packet. Such sophisticated attacks can cause severe consequences, including Man-in-the-Middle (MITM) session hijacking and device impersonation. According to our dataset of 117 real-world CVEs, around 54\% of BLE vulnerabilities are session-based, which forms a huge attack surface (as shown in Figure \ref{fig:vul-stat}, Appendix \ref{appendix:vulsta}). Detecting BLE session-based attacks is challenging, as each packet in the attack sequence is legitimate when examined individually, and can be used in normal BLE connections. A session-based attack is only triggered when these packets are maliciously used by an attacker to form a specific attack sequence.

As a result, if the BLE security mechanism fails to perform inspections on the packet sequence (i.e., at the session level),
it cannot verify the legitimacy of the entire session, thereby becoming prone to session-based attacks.
However, to the best of our knowledge, existing BLE defense frameworks are inefficient against session-based attacks, primarily due to the following two limitations:

\noindent\textit{Only Inspecting Individual Packets.}
%
LBM~\cite{Tian2019LBMAS} leverages stateless filters on individual packets in Linux BlueZ stacks, which can only inspect the packet payloads one at a time. It ignores the forward messages before the target packet and therefore fails to detect session-based attacks without the knowledge of session context.

%

\noindent\textit{Long Patching Window.}
The mainstream countermeasure against session-based attacks is for product vendors to distribute software patches from chip manufacturers. However, the traditional firmware update mechanism has many drawbacks in a fragmented IoT environment, where vendors usually maintain many BLE devices with different chips. For vulnerabilities that affect multiple BLE stacks, such as Specification weaknesses, firmware updates usually result in a long vulnerable window (typically months or years) after vulnerability disclosure. Vendors have to wait for a Specification update and then for each manufacturer to develop a patch. After that, vendors must test it with their custom stacks, recompile the firmware, and update the corresponding user products. In addition, the traditional software update mechanism suffers from disruptive device reboots which are unacceptable for time-sensitive devices, and irreversible installations which exacerbate the risk of faulty patches. In the worst case, many weak IoT devices do not support firmware updates due to limited IO capability, leaving traditional software patches backward incompatible \cite{Wu2020BlueShieldDS}.

Other security frameworks do not support comprehensive BLE traffic filtering and are therefore inadequate for mitigating session-based attacks. BlueShield \cite{Wu2020BlueShieldDS} deploys additional monitor devices to capture cyber-physical features of malicious advertising packets. However, it only defends against spoofing attacks and has to work in a stationary BLE network, overlooking a large number of mobile BLE products including wearable devices, mobile medical devices, etc. 
Other IoT security solutions~\cite{Li2020T2PairSA, Jin2020HarnessingTA,Wang2022ProFactoryII} focus on smart homes and Cloud peripheral protocols and thus fail to study BLE in a fine-grained way. For instance, T2Pair~\cite{Li2020T2PairSA} proposes a Universal Operation Sensing scheme for users to pair wireless devices securely. Unfortunately, many BLE attacks happen beyond the pairing phase.


In this paper, 
we present \textbf{\B} (\textbf{Blue}tooth LE \textbf{S}tate-a\textbf{WA}re securi\textbf{T}y framework), a lightweight security solution for IoT devices, which can mitigate BLE session-based attacks. \B\ records session context and monitors sequential actions of BLE connections. With collected session context, \B\ is capable of performing a stateful inspection on the session level in addition to individual packets.
Specifically, we leverage finite state machine (FSM) to detect session-based attacks. We capture the patterns of attack sequences and model them as state transition paths in FSM. During BLE connections, \B\ updates the FSM based on session context and prevents session-based attacks from transiting along malicious paths. \B\ is highly flexible for vendors to customize at runtime, which supports adding and removing security policies (namely, transition rules of FSM) to extend FSM for new attacks. 

Although the concept of FSM has been involved in Bluetooth security by previous works, \B\ serves distinct motivations and goals, and functions in a different manner. For example, L2Fuzz \cite{l2fuzz} utilized the FSM as a fuzzing component to increase code coverage. It focused on the L2CAP protocol of Bluetooth Classic, while \B\ covers both controller and host layers of BLE. BLEdiff \cite{blediff} abstracted FSM models from different stack implementations to compare and identify deviant non-compliant behaviors as potential vulnerabilities. Profactory \cite{Wang2022ProFactoryII} took user-defined FSMs to automatically generate memory and multiplex safe stack implementations. By contrast, \B\ presents a novel FSM to model the general behavior of BLE standard and various attack patterns, which aims to capture malicious session sequences by monitoring FSM transitions.




\B\ FSM is abstracted from the Specification to monitor BLE connections, which is comprehensive to model various attack patterns. To capture session states and update FSM at runtime, \B\ deploys a compact set of hooks at the Link Layer (LL) and Security Manager Protocol (SMP) of BLE stacks. Hooks at these layers can comprehensively capture session context for modeling various BLE attacks. This is achieved by leveraging the universal packet format of LL and SMP defined by Bluetooth Specification. In BLE connections, LL has access to all session data and SMP controls security procedures such as key sharing and authentication. Therefore, hooks on these two protocols are sufficient to monitor session progress and capture FSM events.


To overcome the disadvantages of the traditional patching mechanism, we develop a lightweight eBPF framework \cite{ebpf} to facilitate \B\ patch distribution. Security patches are first written in C and then compiled into eBPF programs. For a vulnerability that affects multiple devices, vendors can create one common patch and deploy it across all victims. The insights here are (1) BLE stacks of different manufacturers share similar layers and procedures defined by Bluetooth Specification, (2) eBPF bytecode can be executed across different chip architectures. Therefore, vendors can build universal FSM models and transition rules for cross-device deployment, reducing the long vulnerable window after vulnerability disclosure.
Vendors can transmit eBPF programs to victims via BLE and directly integrate them into \B . The updating strategy brings two advantages: (1) It allows vendors to patch vulnerabilities for devices that do not support firmware updates, and (2) It avoids device reboot and firmware recompilation when installing new patches. Such convenience is essential to BLE devices in security and time-sensitive scenarios, such as manufacturing plants, and medical and health care.


We implement \B\ on 5 embedded devices with different chips and BLE stacks to demonstrate its adaptability in heterogeneous IoT environments. 
We systematically investigate various BLE attacks and build an evaluation dataset with 101 real-world vulnerabilities. Our evaluation result shows that \B\ is capable of mitigating 76.1\% (35 / 46) of session-based attacks and 96.4\% (53 / 55) of packet-based attacks, which outperforms existing defense frameworks like LBM \cite{Tian2019LBMAS}. 
In our end-to-end application test, \B\ patches introduce an average of 0.073\% flash overhead and negligible latency, which can be considered controllable in resource-constrained IoT devices.

We summarize our main contributions as follows:

\begin{itemize}
\setlength{\itemindent}{0pt}
\item We design and implement \B\ as the first state-aware security framework against BLE session-based attacks. It performs session-level inspection to monitor sequential actions of BLE connections, in addition to inspecting individual packets. 
\item We develop a lightweight eBPF framework to facilitate patch distribution of \B . Compared with the traditional firmware patching mechanism, \B\ transmits patches as eBPF programs via BLE, avoiding firmware recompilation and device reboot. 
\item For a vulnerability that affects multiple devices, vendors can develop one common eBPF patch and deploy it to all victims, reducing the long vulnerable window after vulnerability disclosure.
\item We implement \B\ on 5 real-world IoT devices with different BLE stacks and chip architectures. On our dataset with 101 real-world vulnerabilities, \B\ can mitigate 76.1\% of session-based attacks and 87.1\% of all attacks. In our end-to-end application evaluation, \B\ patches introduce less than 0.073\% of memory overhead on average and negligible latency.
\end{itemize}

\section{Background and Motivation}
\label{sec:background}

\subsection{BLE Basics}
\label{sec:bluetooth overview}


In a BLE connection, two devices work in a central-peripheral scheme. During different phases of BLE sessions (as shown in Figure \ref{fig:back}, Appendix \ref{sec:connection}), two devices have different names according to the Specification, such as advertiser, scanner, and initiator. For simplicity, we use central and peripheral to distinguish peer BLE devices in this paper.

The BLE protocol stack consists of a controller and a host. The controller contains the physical layer and the logical link layer, which is typically implemented in Bluetooth chips with \M\ customization to match the hardware. The host contains a Logical Link Control Adaptive Protocol (L2CAP) layer and some upper protocols to support application profiles. The Host Controller Interface (HCI) provides a coarse-grained communication channel between the host and the controller. 
Between an HCI command and an HCI event, there may be multiple actions in the lower controller layer. Therefore, unlike LBM, which hooks at HCI, \B\ hooks at the lower link layer for comprehensive inspection of malicious connections. LL hooks capture the session context at runtime, where all raw incoming bytes are decoded for the first time and accessible as plaintext in the stack.



\subsection{{Extended Berkeley Packet Filter (eBPF)}}
\label{subsec:ebpf background}

eBPF is a virtual machine in the Linux kernel that allows users to load userspace code, known as eBPF programs, into kernel space for better interaction with the kernel. 
These programs can be written in C code and compiled using the LLVM toolchain.  After verification, the eBPF program is compiled into native code, ensuring safe and high-performance execution inside the kernel. Because of its extensive and powerful capabilities, eBPF has become a popular choice for filtering network packets~\cite{EBPFCLOUD, DeMarinis2020sysfilterAS,Tian2019LBMAS,Nelson2020SpecificationAV} and implementing complex kernel extensions~\cite{201978,205621,bpfper,Findlay2021BPFContainFT,Findlay2020bpfboxSP,Ghavamnia2020TemporalSC,180196,Song2015ExploitingAP}.

Previous works~\cite{Femto-containers, rapidpatch} demonstrate that eBPF can be ported to run efficiently in resource-constrained embedded devices, outperforming other script runtimes such as MircoPython~\cite{MicroPython}, WASM~\cite{wasm}, RIOTjs~\cite{riotjs}.
%
Since using scripting languages to define firewall policies offers greater flexibility and extensibility than traditional configuration files, it is promising to extend the existing embedded eBPF runtimes to support the BLE state machine and execute policies in the eBPF sandbox. Moreover, since eBPF programs can be separately compiled into bytecode in advance without recompiling the entire firmware, when developers need to add new policies, they can simply transmit the bytecode over BLE or USART which is compatible with millions of legacy embedded devices that may not even support firmware updates. RapidPatch \cite{rapidpatch} has leveraged the scalability of eBPF bytecode on multiple chip architectures to patch across RTOSes. However, it can only patch RTOSes with the same vulnerable libraries (e.g., OpenSSL). By contrast, \B\ extends the patch scalability to different stack implementations based on our insight that BLE stacks of different manufacturers share similar layers and procedures defined by Bluetooth Specification.

\subsection{A Motivating Example}
\label{sec:motivation}

\begin{figure}[t]
    \centering
    \includegraphics[width=1\linewidth, trim=0 0 10 0,clip]{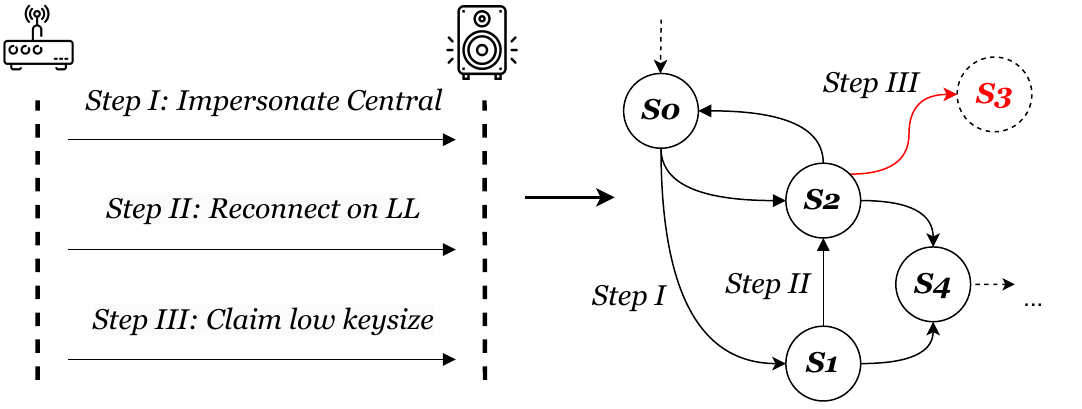}
    \caption{\textbf{BLE KNOB attack \cite{Antonioli2020KeyND} and FSM diagram. The pattern of the KNOB session is modeled as a malicious transition path in FSM. S3 is the exploiting state.}}
    \label{fig:motivation}
\end{figure}

Figure \ref{fig:motivation} presents the famous BLE KNOB attack \cite{Antonioli2020KeyND} as a motivating example. The attacker first jams a benign connection and then launches a MITM attack. It impersonates the central and sends a pairing request with a low Maximum Keysize (7 bytes) to the peripheral, and vice versa. Two victims, who previously used a strong session key (16 bytes), are deceived and downgrade to use a weak encryption key. The attacker then eavesdrops on the ciphertext and brute-forces the low-entropy key, which breaks BLE encryption.

\noindent\textbf{Challenges.}
To defend against KNOB attack, we mainly need to address two challenges in detection and mitigation:

\noindent\textit{(1) To detect the session-based KNOB attack, we need to track the entire attack session rather than individual packets.} Since KNOB is a session-based attack, individual steps of the malicious session are benign. In other words, a peer device claiming low key entropy may not be a KNOB attacker, since weak IoT devices can use short keys for computing resource control. It is impractical to simply reject all pairing requests with low key entropy, which will introduce many false cases and break usability. Therefore, to be compatible with devices that do not support large-entropy keys, the Specification still allows keys smaller than 16 bytes.

\noindent\textit{(2) To mitigate KNOB, a Specification design weakness that affects all BLE devices, we need to address the vulnerable window between attack disclosure and patch update.} For compliant attacks such as KNOB and BlueMirror \cite{Claverie2021BlueMirrorRO}, the existing mitigation is to update the Specification \cite{btsok}. However, for vendors that maintain multiple devices from different \M , distributing Specification updates is extremely time-consuming because they have to wait for all \M\ to develop their patches. It usually takes months or years for \M\ to update BLE stacks with new Specifications and thus leaves a long vulnerable window after vulnerability disclosure. For example, as of January 2024, the BLE stack of Texas Instrument \cite{TIstack} has not yet been updated to Bluetooth v5.4 and supports Advertising Encryption, which was introduced in January 2023 \cite{btspec54}. Even though the Secure Connections Only mode can avoid KNOB, no general-purpose BLE devices are observed to use this mode by default \cite{btsok}. Therefore, most devices are still vulnerable to the KNOB attack. 

We use FSM to model BLE connection actions and capture malicious sequential behaviors. As shown in Figure \ref{fig:motivation}, \B\ captures the pattern of the KNOB session as a malicious transition path in the FSM. Each step of the attack triggers a transition and eventually drives the FSM to the exploiting state, i.e., from S0 via S1 and S2 to S3. A session that performs Step III without I and II will not transit to S3 because it will not enter S2 in the first place. Therefore, by modeling the state transitions of the KNOB session in the FSM, \B\ performs an inspection of the connection context to detect session-based attacks and avoid false cases.
Our eBPF framework allows vendors to develop one common eBPF patch for a compliant attack and deploy it to all victims, reducing the long vulnerable window after the attack is disclosed. We present design details of \B\ in Section \ref{sec:design}.

\section{Threat Model}
\label{sec:threatmodel}


We use the Dolev-Yao model \cite{Dolev1983OnTS} as our threat model, which is compliant with previous works and the Bluetooth Specification. We assume the attacker is in the range of a BLE network that consists of two legitimate devices.
The attacker is capable of eavesdropping on the BLE network to collect all plaintext information publicly broadcast by the victims, such as address, IO capability, and authentication requirements. Also, the attacker can jam the BLE spectrum to block benign connections and inject arbitrary packets into the network. The attacker cannot physically temper the devices. 

We focus on the controller and host part of BLE stacks, which excludes attacks on the physical layer and application layer, such as side-channel analysis \cite{evaluat, linkingble, allowlist} and signal eavesdropping \cite{evaluat, Becker2019TrackingAB}. Based on the observation that BLE connections are in the format of packets, we summarize BLE attacks as two types, namely, \textit{packet-based attack} and \textit{session-based attack}.
A packet-based attack leverages one manipulated packet to exploit the BLE stack, which usually aims at the implementation faults such as bounds check missing and buffer overflow. By contrast, a session-based attack uses a series of crafted packets to construct a malicious attack sequence. In a malicious session, the preceding packets prepare the exploiting conditions for the subsequent packets to eventually compromise the victim.
Note that each step of a session-based attack is legitimate when taken apart and can be commonly used in other benign sessions. A session-based attack is only triggered when these steps form a specific attack sequence. 
%
%

In this paper, we focus on defending against session-based attacks. Besides, \B\ is compatible with previous works in mitigating packet-based attacks by filtering packets. We evaluate \B\ against both session-based and packet-based attacks in Section \ref{sec:secanalysis}. 


\section{System Design}
\label{sec:design}

\subsection{Design Goals and Overview}
\label{sec:goals}

When mitigating BLE session-based attacks in IoT networks in real-time, we are destined to achieve five design goals as follows.

\noindent\textbf{G1: Effectiveness and Extensibility.} 
\B\ should be effective in mitigating known BLE session-based attacks, and easy to extend for new vulnerabilities. 

\noindent\textbf{G2: Compatibility.} 
\B\ should be compatible with heterogeneous chip architectures and BLE protocol stacks in a fragmented IoT environment.

\noindent\textbf{G3: Usability.} 
For vendors, integrating \B\ into BLE stacks and maintaining defense policies should be undemanding. BlueSWAT should be able to compensate for the inadequacy of the traditional firmware update mechanism in complex heterogeneous IoT scenarios, allowing vendors to patch vulnerabilities simply and efficiently.

\noindent\textbf{G4: Comprehensive Mediation.}
\B\ needs to guarantee that all traffic is inspected before handed over to the BLE stack. \B\ should not be bypassed under any circumstances.

\noindent\textbf{G5: Practicality.}
\B\ should only introduce limited performance overhead, and thus be practical on resource-constrained IoT embedded devices.

\begin{figure}[h]
\vspace{-10pt}
    \centering
    \includegraphics[width=1\linewidth, trim=10 0 8 0,clip]{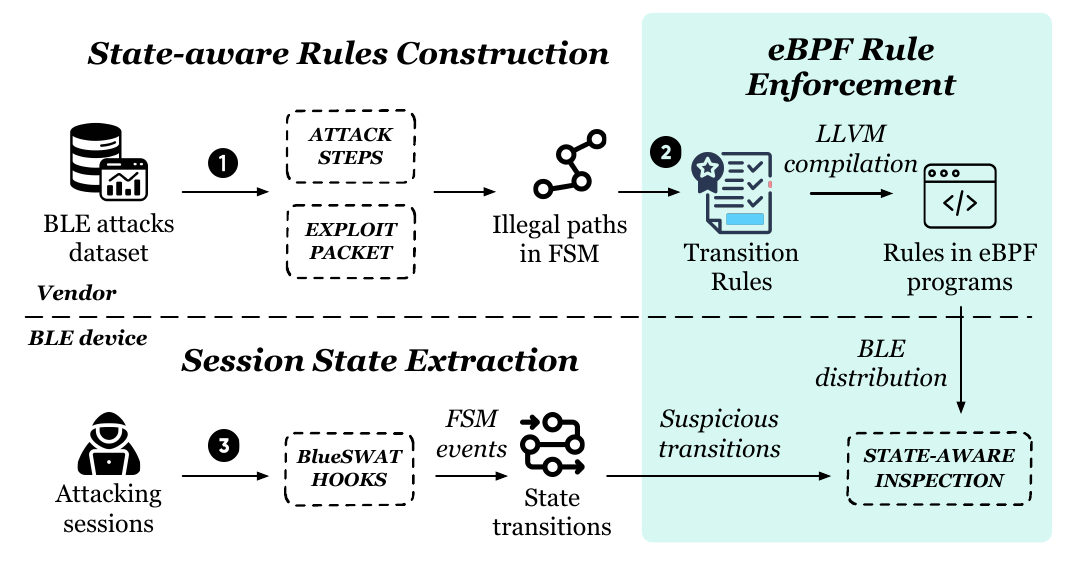}
    \caption{\textbf{Workflow of \B . \ding{182}: Vendors abstract attack patterns and model them as illegal transition paths in FSM. \ding{183}: Vendors compile transition rules into eBPF programs and distribute them to BLE devices. \ding{184}: \B\ captures session events and inspects FSM transitions at runtime.}}
    \label{fig:overview}
\end{figure}

Figure \ref{fig:overview} shows the overall architecture of \B\ that targets to achieve the above five design goals. As BLE connections proceed, \B\ monitors session context in real-time, performing session-level inspection in addition to analyzing individual data packets. Based on the Bluetooth Specification, we build a general FSM to model BLE connections, which can describe sequential actions of different attacks.
Specifically, we capture patterns of session-based attacks, including the attack steps and the final exploit packet, and model them as malicious transition paths in FSM. \B\ prevents FSM from transiting through malicious paths into exploiting states, which indicates connection exploitation. For newly disclosed attacks, vendors can add corresponding transition rules in FSM for inspection. Therefore, we achieve the design goal of Effectiveness and Extensibility (G1).

To overcome the resource limitations of embedded devices and the fragmentation of the IoT environment, we develop a lightweight eBPF framework. It allows developers to write concise C policies for FSM and compile them into eBPF programs. The nature of eBPF bytecode allows \B\ policies to be executed on different chips, which facilitates patch distribution across heterogeneous devices. 
\B\ hooks at the shared layers of different stacks, which are the LL and SMP layers, to capture monitored attack patterns and state transitions in FSM. For a vulnerability that affects multiple different devices, vendors can develop one common patch and deploy it to all victims. Hence, we meet the design goal of Compatibility (G2). 
For patch distribution, vendors can transmit eBPF programs via BLE and directly integrate them into \B , avoiding firmware recompilation and device reboot. Combined with the cross-chip and cross-stack deployment capability, extending \B\ for new attacks becomes more efficient and convenient than the traditional firmware update mechanism. The hooks of \B\ are designed based on the Bluetooth Specification, which maintains a minimal dependency on specific stack implementations. 
Our compact set of hooks requires less than 10 lines of code insertion, which ensures \B\ undemanding for vendors to integrate into both open-source and closed-source firmware. Hence, we achieve the design goal of Usability (G3).

The lower hooks of \B\ are between the link layer and physical layer, where BLE traffic is first decoded and processed by the stack controller. By resolving input session data at the link layer, \B\ performs an inspection on plaintext session packets before they reach higher layers and exploit potential vulnerabilities. Hence, we achieve the design goal of Comprehensive Mediation (G4).
Our evaluation result shows that \B\ introduces a controllable memory footprint and negligible runtime latency, meeting the design goal of practicability (G5).

\begin{figure*}[t]
    \centering
    \includegraphics[width=1\linewidth, trim=15 20 35 38,clip]{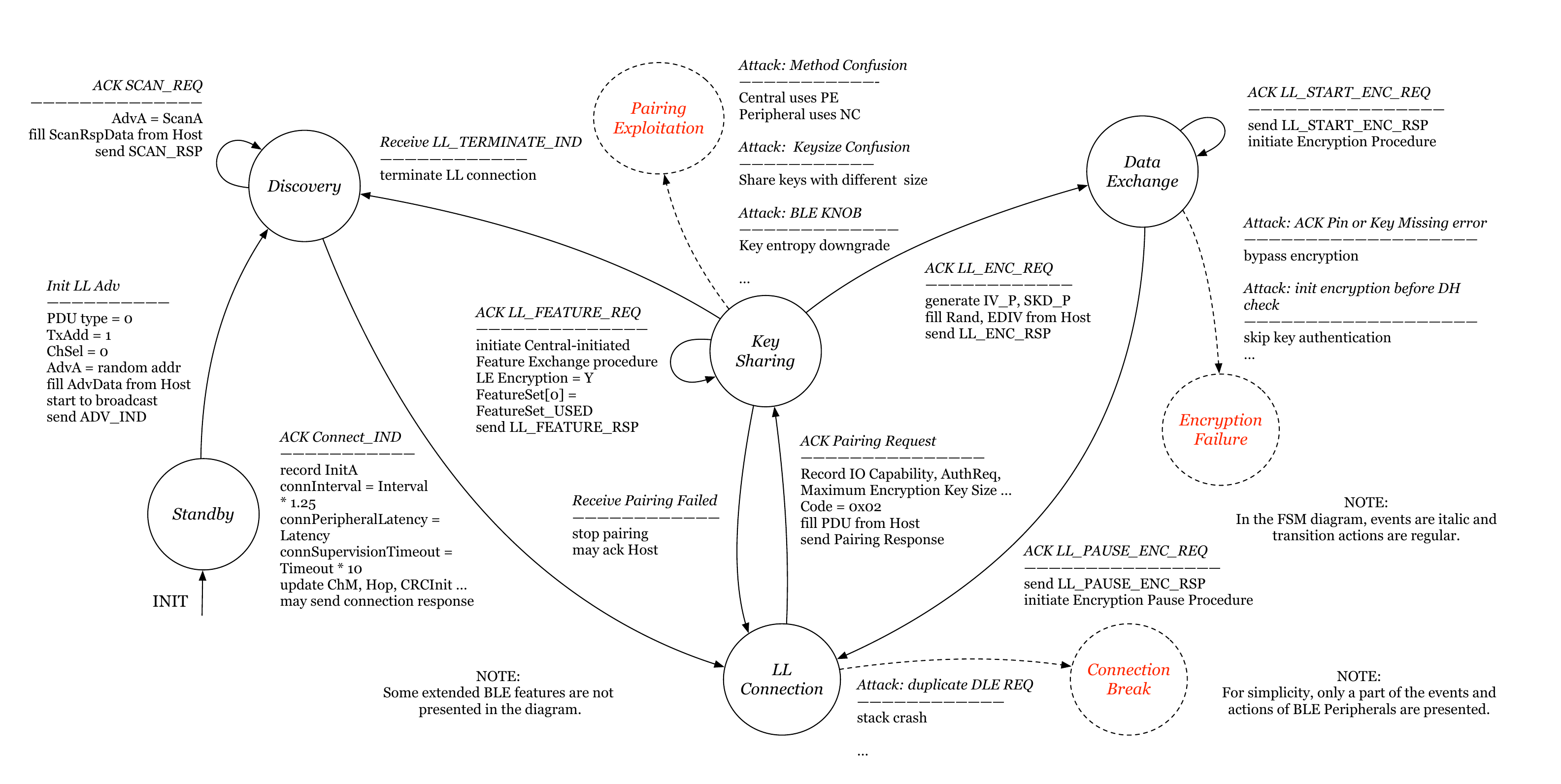}
    \caption{\textbf{\B\ FSM.}}
    \label{fig:fsm}
\end{figure*}

\subsection{State-aware Rules Construction}

Based on the Bluetooth Specification, we build an FSM to model the BLE connection actions and capture BLE attacks. As shown in Figure \ref{fig:fsm}, a session-based attack will transit along a malicious path in FSM, which eventually enters an exploiting state and is rejected by \B\ transition rules.

\subsubsection{States}

The FSM state space consists of benign states and exploiting states. Legal sessions drive FSM in benign states while malicious sessions trigger transitions into exploiting states. The FSM is abstracted from the Bluetooth Specification, which is comprehensive to model the behaviors of different BLE stacks.

\noindent\textbf{Benign States.} Benign states describe the behaviors of legitimate BLE connections.

\textit{Standby.} Standby state is the initiating state. In this state, BLE devices are active but not discoverable. They do not transmit or receive any packets. When a connection is finished or terminated by \B , the BLE stack is reset to the Standby state.

\textit{Discovery.} In this state, centrals scan for advertising peripherals. BLE advertising packets are unencrypted and contain device information such as device address and name. Many attacks leverage public messages in the Discovery state to perform impersonation and spoofing attacks. Peripherals can use the advertising encryption procedure (introduced in Bluetooth v5.4 \cite{btspec54}) with previously paired centrals. But to connect with new peer devices, exchanged messages in the Discovery phase have to be plaintext. After a peripheral response to the connection request from a central, two devices enter the LL Connection state.

\textit{LL Connection.} In this state, an LL connection is established. Connection parameters, such as interval and channel map are exchanged. The messages in the LL Connection state are also plaintext. Hence, many attacks exploit firmware vulnerabilities with crafted malicious packets in this state. After a peripheral response to a pairing request from a central, two devices enter the Key Sharing state and initiate the pairing procedure. 
Two devices may skip pairing and exchange data in plaintext if they do not support LL encryption.

\textit{Key Sharing.} Two devices perform the SMP pairing procedure in this state. They exchange device features and then generate and distribute a shared key. BLE devices can choose multiple pairing methods and association models based on their IO capability and authentication requirements. Most of the specification weaknesses are discovered in this state. They allow attackers to downgrade the security models into vulnerable ones to skip authentication, break encryption, etc. After two devices exchange the shared key, they enter the Data Exchange state.

\textit{Data Exchange.} In this state, two devices establish an encrypted connection and transmit BLE data. They derive a session key from the shared key and then perform the encryption procedure. If the connection is correctly encrypted, BLE data transmission is relatively secure. However, nearly 80\% of existing BLE devices communicate in plaintext \cite{BLEhijack}, which leaves a huge attack surface. After the data transmission is over, the BLE stack returns to the Standby state and waits for future connections.

\noindent\textbf{Exploiting States.} Malicious BLE sessions will eventually drive the FSM into exploiting states and trigger \B\ alerts.

\textit{Discovery Error.} Few attacks compromise the Discovery state because two devices are not connected. In practice, we only observe one vulnerability (i.e., CVE-2017-13211) caused by repeated scan requests. Therefore, the Discovery Error state is not presented in Figure \ref{fig:fsm} for simplicity. Many attacks leverage the plaintext messages in the Discovery state to exploit other connection states.

\textit{Connection Break.} After the LL connection establishment, attackers can invoke some BLE procedures and exploit victim stacks. Since the connection is not encrypted, attackers can inject malicious plaintext packets with invalid fields to exploit implementation faults, which cause buffer overflow, device crash, etc. 

\textit{Pairing Exploitation.} The pairing procedure is a critical security procedure of BLE for authentication and encryption. Hence, many attacks aim to compromise device pairing, such as association model and shared key compromise. 

\textit{Encryption Failure.} If two devices correctly perform encryption, the Data Exchange state is relatively secure. However, some implementation faults of manufacturer stacks allow attackers to cause encryption failures and compromise BLE confidentiality.

\subsubsection{Events.}
\B\ FSM events are compliant with the Specification and comprehensive to model all BLE connections. 
When \B\ captures these events, it automatically updates the FSM to follow the session progress. 
Since BLE centrals and peripherals are identical in security-related protocols and peripherals are more pervasive \cite{blediff}, we describe the events from the perspective of peripherals.

\noindent{(1) \textit{Advertising Packets.}}
When LL starts to broadcast advertising packets such as \texttt{ADV\_IND} and \texttt{SCAN\_REQ} PDU, FSM transits from the Standby to the Discovery state.

\noindent{(2) \textit{Connection Indicators.}} 
In the Discovery state, when the peripheral receives a \texttt{CONNECT\_IND} PDU and responds with a positive Connection Response, FSM transits to the LL Connection state.

\noindent{(3) \textit{Pairing Indicators.}}
If two devices are not bonded, they need to perform the SMP pairing procedure. When the peripheral receives a Pairing Request and responds with a positive Pairing Response, FSM transits from the LL Connection to the Key Sharing state.

\noindent{(4) \textit{Encryption Indicators.}}
In the Key Sharing and the LL Connection state, when the peripheral receives a \texttt{LL\_ENC\_REQ} PDU and responds with a positive \texttt{LL\_ENC\_RSP} PDU, FSM transits to the Data Exchange state.

\noindent{(5) \textit{\B\ Alerts.}}
When a session enters an exploiting state, \B\ sends an alert to FSM and terminates the connection. FSM transits to the Standby state whenever an alert is accepted.

\subsubsection{Variables.}

Apart from tracking session states, \B\ records a few security-related variables used in previous benign connections. The variables help \B\ in recognizing attacks with session history.

\noindent{(1) \textit{Device address.}}
\B\ records MAC addresses to recognize paired devices. If a peer device uses RMA, \B\ records the real address resolved by the stack.

\noindent{(2) \textit{SMP parameters.}}
\B\ records security critical SMP parameters used in benign connections, such as session key entropy, pairing methods, and association models. Together with the device address, \B\ can prevent an impersonator or MITM from downgrading the SMP security level. 

\subsubsection{Malicious Paths.} The sequential actions of session-based attacks are modeled as malicious transition paths in FSM. Malicious paths are driven by attack-specific events and eventually enter the exploiting states. As sessions proceed, \B\ collects monitored states and updates FSM at runtime. A session triggers a \B\ alert when its transition path from the initiating state covers a malicious path. Paths with parts of the malicious paths are considered benign. Therefore, \B\ can avoid false cases while tracking session sequences.

\subsubsection{Transition Rules} Transition rules check whether FSM events are in accord with attack patterns. If an event sequence transits along a malicious FSM path, the monitored session will drive the FSM into an exploiting state and trigger a \B\ alert. The rules are specific to attacks, which can be added after new vulnerability disclosure.

\subsubsection{An Example: BLE KNOB attack} We use the famous BLE KNOB attack \cite{Antonioli2020KeyND} as an example to explain our defense strategy in detail. Specifically, KNOB is represented as a malicious path in FSM with the following sequence of (state, event) pairs:

\vspace{5pt}
\noindent (\textbf{State}: \texttt{Standby}, \textbf{Event}: Device init.)\\
\noindent (\textbf{State}: \texttt{Discovery}, \textbf{Event}: \texttt{ADV\_IND} sent)\\
\noindent (\textbf{State}: \texttt{LL} \texttt{Connection}, \textbf{Event}: \texttt{CONN\_REQ} received and \\ \texttt{CONN\_RSP} sent, , hit a paired record in \B )\\
\noindent (\textbf{State}: \texttt{Key} \texttt{Sharing}, \textbf{Event}: Pairing Request received and Pairing Response sent) \\
\noindent (\textbf{State}: \texttt{Pairing} \texttt{Expoitation}, \textbf{Event}: smaller \texttt{Keysize} than previously used one)
\vspace{5pt}

When the victim pairs with the peer for the first time, \B\ records the peer device identity and the agreed key entropy. In a future reconnection, a KNOB attacker hijacks the session after an LL connection is established. At this time, the FSM has transited from \textit{Standby} through \textit{Discovery} to \textit{LL Connection}. 
When the attacker impersonates the peer and claims a smaller \texttt{Maximum} \texttt{Keysize} in Pairing Request, \B\ compares the new Keysize with the previously used one and detects a key entropy reduction. It violates \B\ transition rules and drives the FSM into the Pairing Exploitation state. Therefore, the KNOB session transits along a malicious path in FSM and will be terminated by \B .

In real-world scenarios, a short key may be necessary if the device is too weak or the BLE service needs more usability than security. In this case, \B\ allows two devices to use a weak key if they pair for the first time. In other words, a session that transits along a part of the malicious path is allowed by \B . Only when a session covers the entire malicious path will it be considered as a session-based attack. In general, mitigation of \B\ does not impact weak devices or add restrictions to the encryption procedure. It only forbids reducing key size rather than using a weak one by default.
Unfortunately, for packet-level defending methods like LBM, they have to drop any Pairing Request packets with field \texttt{Maximum} \texttt{Encryption} \texttt{Key} \texttt{Size} equal to 7. Such methods could severely impact or disable pairing for weak BLE devices that require short session keys. Compared with the mitigation of Bluetooth SIG, our mitigation does not require updating the Specification and breaking backward compatibility.

\subsection{Session State Extraction}
\label{sec:sse}

\B\ deploys a compact set of hooks in BLE stacks to extract session states in real-time. 
The lightweight hooks require less than 10 lines of code modification on BLE stacks, which facilitate the integration of \B\ into both open-source and closed-source embedded firmware.

\subsubsection{\B\ Hooks.}
Although BLE stacks are separately implemented by manufacturers with various solutions, their basic architecture and behaviors are similar, such as BLE packet formats, HCI command formats, pairing procedures, etc. This similarity determined by the Bluetooth Specification allows us to design a set of universal hooks at certain layers of the stacks, regardless of implementation details of \M . Such design significantly reduces the engineering effort required to integrate \B\ with the existing BLE stacks, while providing \B\ policies with the capability of cross-stack defense.

Specifically, \B\ focuses on two security critical protocols in BLE, which are the Link Layer (LL) protocol and the Security Manager Protocol (SMP). LL is responsible for managing BLE connection states and directly interacts with the physical layer. It also controls the encryption and decryption procedure of BLE, which means LL is where plaintext data can be accessed in the stack for the first time. Security Manager Protocol (SMP) serves as the core architecture of BLE security mechanisms, such as device pairing and encryption. Hence, monitoring the behaviors of SMP is crucial for BLE security. Bluetooth Specification stipulates the packet formats for SMP, therefore we can universally resolve the SMP traffic at the entrance. Specifically, the SMP parser hook acquires important session parameters for the FSM models, such as encryption keys and pairing association models, which are broadly manipulated by session-based attacks \cite{Haataja2010TwoPM, Antonioli2020KeyND, Shi2023FormalAA, Antonioli2020BLURtoothEC, Claverie2021BlueMirrorRO, Wu2022FormalMD, Zhang2020BreakingSP, Tschirschnitz2021MethodCA}.

To ensure that all BLE traffic is subject to \B\ inspection, we integrate hooks into the link layer which serves as the entry point for traffic into the BLE stack. 
As shown in Figure \ref{fig:hooks}(a), the LL RX hook resolves the traffic where the packet data is accessible in plaintext for the first time after it is received by the BLE controller. It monitors the session packets and collects the state transitions that need to be verified. 
Simply monitoring the RX path is not sufficient to infer connection states, since the device may require data retransmission to renegotiate parameters or fix error packets. In this case, packets may be accepted by \B\ but rejected by the stack, which should not be allowed to trigger transitions.
Therefore, we design the LL TX hook to monitor the output traffic from the stack, collaborating with the RX hook to determine the states of the sessions. When the stack responds with error messages to indicate retransmission or disconnection, \B\ will not update the new states of the illegal input to FSM, even if they pass the verification. \B\ FSM operates at a higher level than inner FSMs of the stack (e.g., L2CAP FSM). It does not share common states with inner FSMs or temper with their transitions. When a session triggers an event of \B\ FSM, \B\ waits for the inner FSMs to finish transitions and transits along with the response (TX) messages.

\begin{figure}[!h]
\vspace{-3pt}
    \centering
    \begin{subfigure}[b]{1\columnwidth}
        \begin{lstlisting}[language=C]
// nimble/controller/src/ble_ll_conn.c:
void ble_ll_conn_rx_data_pdu(struct os_mbuf *rxpdu, 
struct ble_mbuf_hdr *hdr){
  ...
  if (IFW_DC_LL_CTRL_PARSER(connsm, rxpdu)){
    goto conn_rx_data_pdu_end;
}
  ...
}
        \end{lstlisting}
        \caption{\textbf{LL RX parser for control PDUs.}}
        \label{code:ll_Rx_parser}
    \end{subfigure}
    
    \vspace{15pt}
    \begin{subfigure}[b]{1\columnwidth}
    \begin{lstlisting}[language=C]
// nimble/host/src/ble_sm.c:
int ble_sm_rx(struct ble_l2cap_chan *chan){
  ...
    if (IFW_SMP_PARSER(chan)){
        return BLE_HS_EUNKNOWN;
    }
  ...
}
        \end{lstlisting}
        \caption{\textbf{SMP RX parser.}}
        \label{code:smp_Rx_parser}
    \end{subfigure}
    
    \vspace{10pt}
    \caption{\textbf{\B\ hooks in Mynewt NimBLE.}}
    \label{fig:hooks}
    \vspace{-2pt}
\end{figure}

\subsubsection{State Extraction.}
With the hooks monitoring session traffic, \B\ uses eBPF programs to capture the inspected FSM events based on transition rules. 
Once a state transition is triggered by the captured event, \B\ invokes corresponding transition rules based on the state and event, and performs
a validity check. \B\ policies are mapped to corresponding states and events and are only invoked when target states are changed. This scheme ensures that \B\ introduces less performance overhead in comparison to existing solutions such as LBM, which requires the traversal of all the filters for every packet check.

\subsection{eBPF-based Rule Enforcement}

We develop a lightweight eBPF framework for IoT platforms based on the Linux uBPF library \cite{ubpf}. 
The transition rules are written in C and compiled into eBPF bytecode in advance. 
When a session triggers an FSM transition, \B\ loads the eBPF programs from the policy cache for verification.
Compared with implementing software patches in C firmware, our eBPF framework provides three advantages: (1) eBPF programs can be transmitted via BLE and dynamically loaded by \B. Hence, patch update does not require firmware recompilation and device reboot. (2) eBPF bytecode can be executed across different chips regardless of their architectures, which ensures that \B\ is compatible in the fragmented IoT environment. (3) eBPF programs consume limited memory resources and introduce negligible runtime overhead as shown in Section \ref{sec:eval}, and thus provide \B\ with real-world practicality across resource-constrained devices. 

\subsubsection{Patch Compilation and Transmission.}
We use the bpf LLVM backend to compile C patches into eBPF bytecode.
Instead of deploying new patches through traditional software update mechanisms (which encounter many drawbacks as mentioned in Section ~\ref{sec:intro}), we transport the eBPF bytecode to \B\ via BLE. Specifically, we implement a new BLE service and characteristic for patch transmission. In the service, \B\ stores new patches locally for future execution. With our eBPF framework, updating policies does not need to recompile firmware and reboot devices because eBPF programs can be directly loaded into the kernel for execution. For weak IoT devices that do not support firmware updates, transmitting patches via BLE enables them to accept security updates. 

\subsubsection{Cross Stack and Architecture Deployment.}
Since all different BLE stacks are implemented following the Bluetooth Specification, \B\ is designed to maintain minimal dependency on specific implementation details of the vendor stacks. Specifically, the only part of \B\ that is related to specific stack implementation details are the hooks. For example, LL entries of the NimBLE and Zephyr stacks are different, and so are the data structures storing packet payload. \B\ hooks on the data structures and extracts states for FSM, where states are universal to all BLE stacks. The FSM model and security policies are all in accord with the universal FSM states. Since we implement multiple instruction sets for embedded devices, the eBPF bytecode can be directly executed in eBPF virtual machines across different chips.
Therefore, for a vulnerability that affects all stacks, vendors can develop one patch with our universal states and deploy it to all end devices. 

\subsubsection{Defense Extension.} 
\B\ is highly flexible for vendors to customize, which supports dynamically adding and removing policies. To this end, we introduce \B\ \textit{specifications} which leverage the eBPF map to maintain FSM security policies. Specifically, specifications maintain two sets of (key, value) pairs in the eBPF map: (1) the hooks and the monitored events, and (2) the events and the security policies. Therefore, to modify events and policies, developers only need to change the eBPF map entries with specifications. For example, to remove the transition rule \texttt{MYNEWT\_INTERVAL} on the connection interval value, developers can simply remove the map entry of (interval, \texttt{MYNEWT\_INTERVAL}). Then, \B\ will stop inspecting the interval value and remove the policy. Since specifications are also eBPF programs transmitted to \B\ via BLE, the extension operations of \B\ share the advantages of patch distribution, such as avoiding firmware recompilation and introducing limited performance overhead.

\section{Implementation}
\label{sec:imple}

We implement \B\ on 5 real-world devices with mainstream BLE stacks and architectures as shown in Table \ref{tab:board}. 
The open-source ZephyrOS \cite{zephyr} and Apache Mynewt \cite{mynewt} are popular small-footprint kernels designed for resource-constrained and embedded systems, which have been widely adopted by many commercial products \cite{zephyrproducts}. The other stacks are also broadly used in real-world products, which are produced by famous manufacturers including TI \cite{TI}, Espressif \cite{Espressif} and Bouffalo \cite{Bouffalo}. The evaluation boards run on mainstream IoT architectures, including ARMv7-M of Cortex-M chips, Xtensa of LX6, and RISC-V of BL618. 

We implement \B\ as a kernel extension and install several hooks in the BLE protocol stacks. Due to different manufacturer implementations of the stacks, the number of \B\ hooks slightly changes to meet the demand for comprehensively processing BLE traffic. As shown in Table \ref{tab:hook}, we hook \B\ into 5 different BLE stacks, among which 3 are closed-source. In general, \B\ needs no more than 5 hooks and requires inserting less than 10 lines of code. 
In most cases, we need 2 LL RX hooks and 1 LL TX hooks to fully monitor the input traffic because the stacks separately process LL advertising PDUs and data PDUs with L2CAP payload. 
The reverse engineering process is relatively simple for closed-source stacks (i.e., TI SimpleLink \cite{TIstack} and ESP-IDF \cite{Espidf}) since we only need to locate the LL and SMP entrance functions. We do not need to reverse-engineer the packet payload formats because they are uniform to the Bluetooth Specification.
The compact design of hooks significantly reduces the engineering effort to integrate \B\ into the fragmented IoT environment. 

\begin{table}[h]
\vspace{5pt}
    \centering
    \caption{\textbf{Real-world devices used in evaluation. Stacks with * are partly closed-source.}}
    \label{tab:board}
    \resizebox{\columnwidth}{!}{
    \begin{tabular}{lllll}
    \toprule[0.8pt]
       \textbf{Device}  & \textbf{Manufacturer} & \textbf{Processor} & \textbf{Architecture} & \textbf{BLE Stack} \\
    \specialrule{0.05em}{3pt}{3pt}
        nRF51833 DK & Nordic. & Cortex-M0 & ARMv7-M & NimBLE \\
        CC2640R2 & TI. & Cortex-M3 & ARMv7-M & SimpleLink* \\
        nRF52840 DK & Nordic. & Cortex-M4 & ARMv7-M & Zephyr \\
        ESP32 & Espressif. & Xtensa LX6 & Xtensa & ESP-IDF* \\
        Sipeed M0P & Bouffalo & BL618 & RSIC-V & Bouffalo* \\
    \bottomrule[0.8pt]
    \end{tabular}
    }
    \vspace{-5pt}
\end{table}

In terms of the eBPF framework, we use several functions from Newlib~\cite{newlib} that are developed as a replacement for libc for embedded platforms. Considering that most embedded platforms do not encompass MMU to offer kernel and user space partitioning similar to the Linux kernel, our eBPF maps eliminate the necessity of copying map data from userspace to kernel. Moreover, we provide Python assembly, disassembly, and Clang compilation tools to support developers in testing and verifying their eBPF programs.
We develop a compatible JIT compiler for Cortex-M3+ MCU with the ARM Thumb-2 instruction set. We utilize two 32-bit registers as a singular 64-bit eBPF register to counter the length disparity between Thumb-2 and eBPF instruction sets. Each eBPF instruction is translated into at least two Thumb-2 instructions. For the safety and accuracy of \B , we incorporate the SFI mechanism for VM interpretation and JIT mode. It allows us to verify instruction numbers, behaviors, and loop iteration times while in interpretation mode. It enables us to prevent dubious C function calls and malicious out-of-bound memory writing within JIT mode as well.

\begin{table}[h]
\vspace{5pt}
\caption{\textbf{Hooks and lines of code inserted into the stacks.}}
\label{tab:hook}
\resizebox{0.77\columnwidth}{!}{%
\begin{tabular}{lccc}
\toprule[0.8pt]
\textbf{BLE Stack} & \textbf{\# LL Hooks} & \textbf{\# SMP Hooks} & \textbf{LoC Inserted} \\
\specialrule{0.05em}{3pt}{3pt}
NimBLE & 3 & 2 & 8 \\
SimpleLink & 3 & 2 & 8 \\
Zephyr & 3 & 3 & 9 \\
ESP-IDF & 3 & 2 & 8 \\
Bouffalo & 2 & 2 & 6 \\
\bottomrule[0.8pt]
\end{tabular}%
}
\vspace{-5pt}
\end{table}


Changes were made to the existing eBPF VM implementations~\cite{rapidpatch, Femto-containers} on embedded devices, adding approximately 1.2k LOC for state machine augmentation. Overall, \B\ encompasses around 2k lines of C code and 1k lines of Python code.

\section{Security Analysis}
\label{sec:secanalysis}

\begin{table*}[t]
\vspace{5pt}
\caption{\textbf{Defense effectiveness of \B\ on session-based BLE attacks. Attacks are grouped into cause and impact. Chip: Affecting multiple chip architectures. Stack: Affecting multiple BLE stacks. D: Discovery, LL: LL Connection, KS: Key Sharing, DE: Data Exchange.}}
\label{tab:defeff}
\resizebox{\textwidth}{!}{%
\begin{tabular}{p{1.5cm}llcclccccc}
\toprule[1pt]
\multirow{2}{*}{\textbf{Category}} & \multirow{2}{*}{\textbf{Impact}} & \multirow{2}{*}{\textbf{Vulnerability}} & \multicolumn{2}{c}{\textbf{Cross-device}} & \makecell[c]{\multirow{2}{*}{\textbf{Monitered Attack Patterns}}} & \multicolumn{4}{c}{\textbf{Affected States}} & \multirow{2}{*}{\textbf{Mitigation}} \\
\cline{7-10} \cline{4-5} 
&  &  & Chip & Stack&  & D & LL & KS & DE &  \\
\specialrule{0.05em}{1pt}{2pt}
\multirow{10}{*}{\makecell[l]{Design\\ Flaw}} & \multirow{7}{*}{Pairing Compromise} & BLE KNOB \cite{Antonioli2020KeyND} & $\blacksquare$ & $\blacksquare$ & Claim smaller key entropy &  &  & \CIRCLE & \LEFTcircle & \multirow{10}{*}{12 / 15 (80.0\%)} \\
& & BlueMirror \cite{Claverie2021BlueMirrorRO} & $\blacksquare$ & $\blacksquare$ & \textit{BLE-A}: use legacy pairing \& reflect commitment scheme &  &  & \CIRCLE & \LEFTcircle & \\
&  & & $\blacksquare$ & $\blacksquare$ & \textit{PE-A1}: force PE \& reﬂect Key Exchange \& reflect Auth-1\&2 &  &  & \CIRCLE &  & \\
&  & & $\blacksquare$ & $\blacksquare$ & \textit{PE-A2}: force PE \& reﬂect Key Exchange &  &  & \multirow{2}{*}{\CIRCLE} &  & \\
&  & & $\blacksquare$ & $\blacksquare$ & \& reflect Auth-1 \& initiate Auth-2 &  &  &  &  & \\
&  & Keysize Confusion \cite{Shi2023FormalAA} & $\blacksquare$ & $\blacksquare$ & Claim different Keysize from history bonds &  &  & \CIRCLE & \LEFTcircle & \\ [2pt]

\cline{2-10}
\specialrule{0em}{1pt}{1pt}

& \multirow{4}{*}{Illegal Service Access} & BLESA \cite{Wu2020BLESASA} & $\blacksquare$ & $\blacksquare$ & \textit{Reactive}: downgrade to no authentication and encryption &  &  & \CIRCLE & \CIRCLE & \\
&  & Downgrade Attacks \cite{Zhang2020BreakingSP}  & $\blacksquare$ & $\blacksquare$ & \textit{Attack I,II,V}: "Pin or Key Missing" error \& R/W attributes &  & \LEFTcircle & \CIRCLE & \CIRCLE & \\
&  &  & $\blacksquare$ & $\blacksquare$ & \textit{Attack III,IV,VI}: "Pin or Key Missing" error &  & \multirow{2}{*}{\LEFTcircle} & \multirow{2}{*}{\CIRCLE} & \multirow{2}{*}{\CIRCLE} & \\
&  &  & $\blacksquare$ & $\blacksquare$ & \& IO capability downgrade &  &  &  & \\
\specialrule{0.05em}{1pt}{2pt}

\multirow{19}{*}{\makecell[l]{Function\\ Error}} & \multirow{7}{*}{Authentication Bypass} & CVE-2022-45190 &  & & Bypass passkey entry in legacy pairing &  &  & \CIRCLE &  & \\
&  & CVE-2020-12860 &  & & Role switch \& access identification &  & \LEFTcircle & \CIRCLE &  & \multirow{19}{*}{19 / 23 (82.6\%)}\\ 
&  & CVE-2023-34625 &  & & Duplicate unlocking messages &  &  & \CIRCLE & \CIRCLE & \\
&  & CVE-2018-16242 & $\blacksquare$ & & Replay ciphertext based on predictable nonce &  &  &  & \CIRCLE &\\ 
&  & CVE-2023-26979 & $\blacksquare$ & $\blacksquare$ & Overwrite Privacy Flag and Reconnection handles &  & \CIRCLE &  &  &\\
&  & CVE-2019-13953 &  & & Specific exploiting commands &  & \CIRCLE & \CIRCLE &  & \\
&  & CVE-2019-12500 & $\blacksquare$ & & "suddenly accelerate" commands &  & \CIRCLE & \LEFTcircle & \CIRCLE & \\ [2pt]

\cline{2-10}
\specialrule{0em}{1pt}{1pt}

& \multirow{4}{*}{Key Compromise} & CVE-2020-16630 & $\blacksquare$ & & Use Just Works \& device impersonation & \LEFTcircle &  & \CIRCLE & & \\
&  & CVE-2019-19192 & $\blacksquare$ &  & Sequential Attribute Protocol requests &  &  & \CIRCLE & &\\
&  & CVE-2019-17520 &  & & Sequential SM Public Key packets &  &  & \CIRCLE &  & \\ 
&  & CVE-2021-3436 & $\blacksquare$ & $\blacksquare$ & Maliciously pair \& Overwrite existing bond &  &  & \CIRCLE &  & \\ [2pt]

\cline{2-10}
\specialrule{0em}{1pt}{1pt}

& \multirow{2}{*}{Encryption Failure} & CVE-2019-19194 & $\blacksquare$ & & Out of order LL\textunderscore ENC\textunderscore REQ \& zero LTK &  & \CIRCLE & \CIRCLE &  & \\
&  & CVE-2020-13593 & $\blacksquare$ &  & Link Layer encryption setup before DH checking &  & \CIRCLE & \CIRCLE & \LEFTcircle & \\ [2pt]

\cline{2-10}
\specialrule{0em}{1pt}{1pt}

& \multirow{6}{*}{Denial of Service} & CVE-2020-10068 & $\blacksquare$ & $\blacksquare$ & Duplicate DLE request packets &  & \CIRCLE &  &  & \\
&  & CVE-2021-3430 & $\blacksquare$ & $\blacksquare$ & Duplicate LL\textunderscore CONNECTION\textunderscore PARAM\textunderscore REQ &  & \CIRCLE &  &  & \\
&  & CVE-2021-3431 & $\blacksquare$ & $\blacksquare$ & Duplicate LL\textunderscore FEATURE\textunderscore REQ &  & \CIRCLE &  &  &  \\
&  & CVE-2018-20957 & &  & Replay pairing requests &  & \CIRCLE & \CIRCLE &  & \\
&  & CVE-2020-27269 & $\blacksquare$ & & Replay communication sequences &  & \CIRCLE & \CIRCLE & \LEFTcircle & \\
&  & CVE-2017-13211 & $\blacksquare$ &  & Repeated BLE scan results & \CIRCLE &  &  &  & \\

\specialrule{0.05em}{1pt}{2pt}

\multirow{4}{*}{\makecell[l]{Runtime\\ Error}} & 
Bounds Check Missing & CVE-2019-16518 & &  & Illegal large power or voltage values &  &  &  & \CIRCLE & \multirow{5}{*}{4 / 8 (50.0\%)} \\ [2pt]

\cline{2-10}
\specialrule{0em}{1pt}{1pt}

& Buffer Overflow & CVE-2023-23609 & $\blacksquare$ & $\blacksquare$ & Sequential L2CAP packets &  & \CIRCLE & \LEFTcircle & \CIRCLE & \\ [2pt]

\cline{2-10}
\specialrule{0em}{1pt}{1pt}

& \multirow{2}{*}{Logic Error} & CVE-2022-2993 & $\blacksquare$ & $\blacksquare$ & Pairing \& Not qualified SMP keys &  & \CIRCLE & \CIRCLE &  & \\
& & CVE-2020-13595 & &  & Discovery \& wrong return numbers of completed packets & \CIRCLE & \LEFTcircle & \LEFTcircle & & \\

\bottomrule[1pt]
\end{tabular}%
}

\flushleft{
\scriptsize{$\blacksquare$}: \footnotesize{Affecting range of an attack.}
\scriptsize{\CIRCLE}: \footnotesize{Target state of an attack.}
\scriptsize{\LEFTcircle}: \footnotesize{The state that is not targeted by an attack but is affected at the same time.}
}
\vspace{-5pt}
\end{table*}

In this section, we evaluate the defense effectiveness of \B\ on real-world attacks and compare it with previous frameworks.

\noindent\textbf{Dataset.}
We systematically collect 117 BLE vulnerabilities in the dataset by November 2023 (as shown in Figure \ref{fig:vul-stat}, Appendix \ref{appendix:vulsta}). Around 54\% of them are session-based, which is left unstudied by previous research. Specifically, we investigate a wide range of vulnerabilities which include 94 CVEs marked with "ble" and "bluetooth low energy" from the CVE database \cite{CVE-base}, 16 documented CVEs of open-source BLE stacks \cite{zephyrbt, nimble}, and vulnerabilities disclosed by previous publications \cite{Sivakumaran2018ASO, Xie2023AccessYT, Wu2020BLESASA, Claverie2021BlueMirrorRO, Zhang2020BreakingSP, linkingble, Antonioli2020KeyND, Shi2023FormalAA, Tschirschnitz2021MethodCA, Garbelini2020SweynToothUM, allowlist}.
In the testing dataset, we exclude 16 records for lacking details and threat model mismatch. Specifically, attacks \cite{Koh2022BLAPBL, allowlist, linkingble, Cayre2021InjectaBLEIM, Xie2023AccessYT} on the BLE physical layer and application layer, such as signal injection and device tracking, are excluded from the testing dataset. In general, the testing dataset contains 101 real-world records, including 46 session-based and 55 packet-based vulnerabilities. We do not claim that our dataset includes all BLE session-based and packet-based attacks, as there may be vulnerabilities that are not publicly studied or assigned CVE records.

\subsection{Defense Effectiveness}
\label{sec:defeff}

Table \ref{tab:eff} shows the defense effectiveness of \B\ and LBM on our CVE dataset with 101 real-world vulnerabilities. We categorize the vulnerabilities into \textit{Design Flaw} of the Bluetooth Specification and implementation faults of vendor stacks, which include \textit{Function Error} and \textit{Runtime Error}.

\B\ can successfully mitigate 87.1\% of them, including 76.1\% of session-based and 96.4\% of packet-based attacks, which outperforms LBM. Since alarms are only triggered when the FSM transits into malicious states, indicating session exploitation, \B\ introduces no FP on the defendable vulnerabilities. For new attacks, vendors can develop new patches with minimal engineering effort and use \B\ eBPF verifier to check policy correctness and eliminate FN. The unresolvable vulnerabilities are mainly implementation faults of the BLE stacks, such as using weak authentication and no encryption. These problems cannot be fixed unless the manufacturers correctly implement their stacks.

\begin{table*}[t]
\vspace{5pt}
\centering
\caption{\textbf{Vulnerabilities that \B\ cannot mitigate.}}
\label{tab:unre}
\resizebox{\textwidth}{!}{%
\begin{tabular}{llccll}
\toprule[0.8pt]

\multirow{2}{*}{\textbf{Category}} & \multirow{2}{*}{\textbf{Vulnerability}} & \multicolumn{2}{c}{\textbf{State}} & \makecell[c]{\multirow{2}{*}{\textbf{Description}}} & \multirow{2}{*}{\textbf{Why \B\ cannot fully mitigate}} \\
\cline{3-4}
& & S & P & & \\

\specialrule{0.05em}{3pt}{3pt}

\multirow{2}{*}{\makecell[l]{Design\\ Flaw}} & BLESA Proactive \cite{Wu2020BLESASA} & \faCheck & & Reconnection fault. & Cannot fix reconnection procedure. \\
& Downgrade VII, VIII \cite{Zhang2020BreakingSP} & \faCheck & & Misused attribute permissions. & Cannot modify attribute settings. \\ 

\specialrule{0.05em}{1pt}{2pt}

\multirow{4}{*}{\makecell[l]{Function\\ Error}} 
& CVE-2020-12730 & \faCheck & & Lack BLE encryption. & Cannot implement encryption mechanism. \\
& CVE-2020-27276 & \faCheck & & Lack of authentication on the communicating entities. & Cannot implement authentication procedure. \\
& CVE-2020-27270 & \faCheck & & Lack of adequate measures to protect encryption keys in transit. & Cannot enforce channel protection. \\
& CVE-2017-18642 & \faCheck & & Receive RGB parameters over cleartext. & Cannot implement encryption mechanism. \\

\specialrule{0.05em}{1pt}{2pt}

\multirow{6}{*}{\makecell[l]{Runtime\\ Error}} & CVE-2018-10825 & & \faCheck & No authentication and encryption. & Cannot implement authentication and encryption. \\
& CVE-2017-17436 & & \faCheck & No encryption. & Cannot implement encryption mechanism. \\
& CVE-2017-17435 & \faCheck & & Fake authentication. & Cannot implement authentication mechanism. \\ 
& CVE-2020-27264 & \faCheck & & Use deterministic keys. & Cannot fix usage of static keys. \\
& CVE-2018-20958 & \faCheck & & Rely on public information for private operations. & Cannot hide MAC address. \\
& CVE-2020-11957 & \faCheck & & Low entropy key. & Cannot fix the key generation procedure. \\

\bottomrule[0.8pt]
\end{tabular}
}
\flushleft{
\footnotesize{S: Session-based vulnerabilities. P: Packet-based vulnerabilities.}
}
\vspace{-5pt}
\end{table*}

\subsubsection{Session-based Attacks Mitigation.}

Among the 46 session-based CVEs in our dataset, \B\ can effectively mitigate 35 of them. We present the details of defendable session-based attacks in Table \ref{tab:defeff}. The attacks on Specification weaknesses mostly target the BLE pairing procedure to compromise the pairing association models by MITM and impersonation.
\B\ successfully mitigates 12 out of 15 Design Flaws that have unique attack patterns. For example, Downgrade attacks \cite{Zhang2020BreakingSP} always leverage a "Pin or Key Missing" error and a read/write attribute to downgrade the session into using plaintext. We add transition rules to the Key Sharing state, which capture corresponding malicious session patterns and transit FSM to the Pairing Exploitation state. To prevent a BLESA attack, \B\ records the security level of the vulnerable attribute used in previous connections. If a remote device with the address of the server downgrades the security level, \B\ will reject the reconnection, discard the shared key, and perform the pairing procedure again. 

In summary, by monitoring the sequential actions with FSM, \B\ can detect session-based attacks based on their corresponding connection patterns. CVE-2019-19194 \cite{19194} and CVE-2020-13595 \cite{13595} initiate BLE procedures in unexpected orders that maliciously compromise the encryption. With the FSM regulating connection progress, \B\ does not allow out-of-order session behaviors because the transition from the Key Sharing state to the Data Exchange state is not allowed except for no encryption.
For replay attacks, \B\ can set a threshold for repeating request and response messages based on specific attack patterns. Repetitive and reflected messages are rarely seen in our observation of benign real-world BLE sessions, although they are not explicitly forbidden by the Specification. For attacks with sequential malicious packets, such as CVE-2019-19192 \cite{19192} and CVE-2020-17520 \cite{17520}, the attacking sessions transit within one benign state and eventually enter exploiting states when they violate corresponding transition rules. 
%
Compared to packet-based attacks which we inspect payload fields, preventing session-based attacks needs to model attack patterns as transition paths in FSM and check against predefined transition rules. 
Caching session states and running additional user-defined rules for verification may invoke performance overhead, but our eBPF framework ensures the introduced memory consumption and runtime latency at a limited level as shown in Section~\ref{sec:eval}.

\subsubsection{Packet-based Attacks Mitigation.}
Among the 55 packet-based CVEs in our dataset, \B\ can effectively mitigate 53 of them as shown in Table \ref{tab:eff}. Especially, it successfully defends against all BLE bounds check missing and buffer overflow errors, which are the most common vulnerabilities of embedded firmware. Most of these errors are packet-based because they can be triggered by one packet with a malformed field. To defend against these attacks, \B\ can easily filter the potential invalid or overflow fields and drop illegal packets. For example, CVE-2021-3432 \cite{3432} triggers a division by zero error when the \texttt{Interval} field of \texttt{CONNECT\textunderscore IND} PDU is set to 0. \B\ defends against it by applying a sanity check on the \texttt{Interval} field and drops malformed packets. For the other packet-based vulnerabilities, \B\ can add the specific exploiting packet payload in security policies and block them when monitoring session context. 
Compared with LBM which can only filter Bluetooth HCI and L2CAP packets, \B\ extends the defense range to BLE controller which provides additional cover for link layer attacks \cite{10069, cve202010061, 3432, 3433}.

\subsubsection{Insufficient Fix.}
There are 13 vulnerabilities in our dataset that \B\ cannot fully mitigate, and we present detailed descriptions in Table \ref{tab:unre}.
The main reason is that these CVEs result from incorrect implementations of the Bluetooth Specification which directly makes BLE functionality not-compliant or incomplete. Specifically, 10 of them are caused by using weak authentication or no authentication and no encryption on sessions. Under the circumstances, \B\ cannot implement authentication and encryption procedures for the stacks. 3 vulnerabilities are caused by improper reconnection handling and attribute permissions misuse of BLE services. The only way to fix these functionality errors is for manufacturers to correctly implement their stacks and replace the buggy firmware. 

\subsection{Comparison With LBM}
\label{sec:comparison}

We compare the defense effectiveness of \B\ with LBM against both session-based and packet-based attacks. As shown in Table \ref{tab:eff}, \B\ outperforms LBM in defending against both session-based and packet-based attacks. LBM is vulnerable to all session-based attacks because it cannot capture the sequential patterns of packet sequences. For example, LBM cannot mitigate CVE-2019-19194 and CVE-2020-13593 because it cannot detect out-of-order malicious sequences. Without recording the history of previous connections, LBM cannot detect BLE KNOB because it does not know the previously used key entropy value. When a reduced key entropy or regeneration request is received, LBM cannot compare it to previous ones and considers it a legitimate request if the packet payload is benign. For replay attacks that send duplicate messages, LBM considers them legitimate because it is not aware of the sequential actions in connections. It individually inspects each message and thus can be deceived by replay attacks. Note that \B\ rules operate on FSMs, examining both packet fields and FSM state transitions, while LBM only checks packet fields. Therefore, to mitigate session-based attacks, LBM has to define rules to drop every packet of one kind which will result in a high FP, or even disable normal BLE functions.

\begin{table}[h]
\vspace{5pt}
\centering
\caption{\textbf{Comparison of \B\ and LBM for mitigating real-world BLE vulnerabilities in our dataset. }}
\label{tab:eff}
\resizebox{\columnwidth}{!}{%
\begin{tabular}{llcccc}
\toprule[1pt]
\multirow{2}{*}{\textbf{Category}} & \multirow{2}{*}{\textbf{Impact}} & \multicolumn{2}{c}{\textbf{LBM}} & \multicolumn{2}{c}{\textbf{BlueSWAT}} \\ \cline{3-6}
 & & S & P &  S & P \\
\specialrule{0.05em}{1pt}{2pt}
\multirow{2}{*}{Design Flaw} & Pairing Compromise & 0 / 5 & 0 / 0  & 5 / 5 & 0 / 0  \\
& Illegal Service Access & 0 / 10 & 0 / 0 & 7 / 10 & 0 / 0 \\

\specialrule{0.05em}{1pt}{2pt}

\multirow{4}{*}{\makecell[l]{Function\\ Error}} & Authentication Bypass & 0 / 10 & 1 / 2 & 7 / 10 & 2 / 2  \\
& Key Compromise & 0 / 4 & 0 / 1 & 4 / 4 & 1 / 1  \\
& Encryption Failure & 0 / 3 & 0 / 1 &  2 / 3 & 1 / 1 \\
& Denial of Service & 0 / 6 & 0 / 0 & 6 / 6 & 0 / 0 \\

\specialrule{0.05em}{1pt}{2pt}

\multirow{3}{*}{\makecell[l]{Runtime\\ Error}} & Bounds Check Missing & 0 / 1 & 15 / 24 & 1 / 1 & 24 / 24  \\
& Buffer Overflow &  0 / 1 & 9 / 18 & 1 / 1 & 18 / 18  \\
& Logic Error &  0 / 6 & 5 / 9 & 2 / 6 & 7 / 9  \\

\specialrule{0.05em}{1pt}{2pt}

Overall & - & 0 / 46 & 31 / 55 & 35 / 46 & 53 / 55 \\
Proportion & - & 0 & 56.4\% & 76.1\% & 96.4\% \\

\bottomrule[1pt]
\end{tabular}%
}
\flushleft{
\footnotesize{S: Session-based vulnerabilities. P: Packet-based vulnerabilities.}
}
\vspace{-5pt}
\end{table}

In addition, LBM can only mitigate 56.4\% of packet-based attacks because it only inspects L2CAP and HCI packet payloads, ignoring the lower controller part of BLE stacks. We note that many attacks (e.g., CVE-2020-10069, CVE-2021-3433, and CVE-2021-3581) exploit LL implementation faults because it is easy to craft plaintext LL packets. Since the pairing and encryption procedures of BLE security are not initiated in the early stages of connections, the LL is completely exposed to a huge attack surface. By contrast, \B\ places hooks in the LL to inspect lower-layer packet payloads. All data to the upper protocols is packed into the LL PDUs, allowing \B\ to universally monitor all session traffic at the entrance of the BLE stacks. We present some descriptions of the packet-based mitigation comparison in Table \ref{tab:pktresult}, Appendix \ref{appendix:packet-based}.

In general, LBM can only mitigate 30.7\% of all vulnerabilities in our real-world dataset, including 56.4\% of packet-based vulnerabilities and none of session-based vulnerabilities. The defense effectiveness of \B\ significantly outperforms LBM. \B\ can successfully defend against 87.1\% of all attacks in our dataset, which includes 76.1\% of session-based attacks and 96.4\% of packet-based attacks.

\subsection{Defense Scalability}
\label{sec:defscale}

We articulate the process of deploying \B\ for new devices and its adaptability on various stacks and architectures. We manually inject 3 vulnerabilities (i.e., CVE-2020-10069, CVE-2021-3430, and BLE KNOB) in the 5 evaluation boards described in Table \ref{tab:board}. 

Firstly, we add \B\ as a component to the kernels and insert the hooks into the stacks. New components can be added to the firmware by modifying a few lines of the project compilation configurations and hooking the stacks requires minimal engineering efforts as demonstrated in Section \ref{sec:imple}. Secondly, we analyze the sequential patterns of the attacks in terms of FSM events and transition paths. Based on the attack patterns, we write C filter rules and use \B\ compilation tools to compile them into eBPF programs. Then, the eBPF programs are uploaded to the vulnerable devices via our BLE services. The eBPF verifier ensures the new programs do not introduce new vulnerabilities. Finally, we test the effectiveness of the policies by reproducing the attacks and observing the attacking result.

In the end-to-end evaluation, the same set of eBPF security policies successfully functions on different processors and BLE stacks. The cross-architecture adaptability of \B\ is achieved by packaging defense policies in eBPF bytecode, which can universally run on ARMv7-M, RSIC-V, and Xtensa SoCs. The cross-stack deployment is achieved by our universal hooks which block stack implementation details and capture universal session states for policies to inspect. 

In a nutshell, with minimal engineering efforts, \B\ can be easily integrated into vendor devices, tested on targeting attacks, and extended across different stacks and chip architectures.
Intuitively, vendors need to ensure the policy \textit{effectiveness} (i.e., the policy actually works on the targeting attack). Meanwhile, \B\ eBPF mechanism can automatically check the policy \textit{correctness} (i.e., the policy does not introduce new vulnerabilities).

\section{Performance Evaluations}
\label{sec:eval}

In this section, we perform an evaluation of the runtime performance of \B , including memory consumption, runtime overhead and power performance on real-world applications. 
We conduct performance evaluations with the Zephyr stack \cite{zephyrbt} on a Nordic nRF52840 DK \cite{dk}, which has a Cortex-M4 SoC running at 64 MHz, 1 MB Flash, and 256 KB SRAM. We build the firmware, compile transition rules into eBPF bytecode, and run Python scripts on Ubuntu 20.04.
We further use Nordic Connect for Desktop~\cite{nrfcon} to customize the nRF52840 dongle and evaluate two real-world Bluetooth applications. 

\subsection{Memory Consumption}



\begin{figure}[!h]
    \centering
    \includegraphics[width=\linewidth, trim=5 0 0 0,clip]{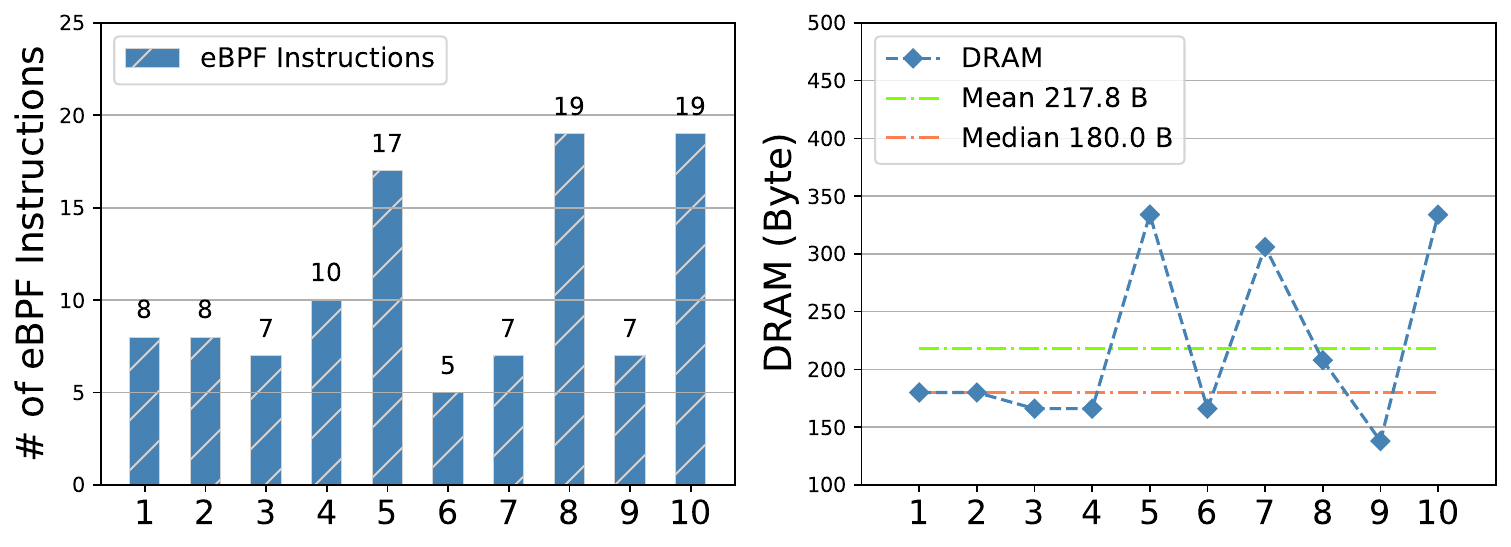}
    \caption{\textbf{Number of eBPF instructions and dynamic memory (DRAM) consumption (Byte) of 10 different rules when JIT enabled.}}
    \label{fig:dram}
\end{figure}

We use a bare BLE peripheral sample as the experiment baseline. It is provided by the Zephyr project as a barebone for vendors to develop their BLE applications, which takes 182.6 KB Flash and 43.1 KB SRAM with a kernel and bare BLE stack. Compared to real-world IoT firmware, the BLE baseline has a smaller size because it does not employ other components such as LED and display. 
When \B\ is integrated with no inspection rules, it takes around 12.8 KB Flash memory and 720 B SRAM. The eBPF framework (which includes a VM environment, JIT compiler, verifier, etc.) takes up most of the memory consumption and the FSM is relatively small. 
When \B\ is integrated with 10 eBPF programs, the size of the firmware increases by 1372 B in Flash, which brings a 0.75\% overhead. On average, one eBPF program takes around 137.2 bytes and brings an overhead of around 0.073\%.
%
%
For eBPF program execution, \B\ needs to allocate additional dynamic memory space for code interpretation. When JIT is disabled, \B\ allocates 24 B to VM for each program. When JIT is enabled, \B\ allocates an additional 217.8 B on average for each program as shown in Figure~\ref{fig:dram}. Note that the dynamic memory is allocated when the program is executed and destroyed when finished. Hence, the dynamic memory consumption is controllable even if more programs are further added.

On average, one ePBF program takes up 137.2 B Flash memory (less than 0.08\% overhead) and 217.8 B dynamic memory, which can be considered controllable. 

\subsection{Runtime Latency}

\subsubsection{Micro benchmark.}
We first load one rule (conn\_chan\_map) in \B\ and generate 1k packets on the Bluetooth RX path to evaluate runtime latency. For the scalability test, we further load 10 different rules in \B\ and generate 1k packets for evaluation. Both benchmarks are tested under VM interpretation mode and Just-In-Time (JIT) mode. 
As shown in Table~\ref{tab:timetab}, when one rule is loaded, the average latency is 1.219 $\mu s$ (78 MCU cycles) under interpretation mode and 1.172 $\mu s$ (75 MCU cycles) under JIT mode. The latency variance is 0.44 under interpretation mode and drops to 0.31 when JIT is on. The maximum overhead is controlled under 5 $\mu s$. In the scalability test, we add 10 different rules to \B\ and the average latency in interpretation mode slightly rises to 1.266 $\mu s$ (81 MCU cycles) while with JIT it drops to 1.094 $\mu s$ (70 MCU cycles). In summary, \B\ introduces negligible latency and maintains the runtime performance when scaling up.

\vspace{5pt}
\begin{table}[h]
\caption{\textbf{The runtime latency (us) introduced by \B .}}
\label{tab:timetab}
\resizebox{0.8\columnwidth}{!}{%
\begin{tabular}{lccccc}
\toprule[0.8pt]
\multirow{2}{*}{\textbf{Benchmark}} & \multicolumn{4}{c}{\textbf{Runtime Latency (us)}} & \multirow{2}{*}{\textbf{Var.}} \\
                                    & Min.     & Max.     & Med.     & \textbf{Avg.}    &                                \\
\specialrule{0.05em}{3pt}{3pt}
\B  -1                          & 0.844    & 4.969    & 1.031    & \textbf{1.219}   & 0.44                           \\
\B -1-JIT                      & 1.031    & 4.953    & 1.031    & \textbf{1.172}   & 0.31                           \\
\B -10                         & 0.906    & 4.891    & 0.922    & \textbf{1.266}   & 0.89                           \\
\B -10-JIT                     & 0.906    & 4.984    & 0.922    & \textbf{1.094}   & 0.45                           \\
\bottomrule[0.8pt]
\end{tabular}%
}
\end{table}
\vspace{-5pt}

\begin{figure}[h]
    \centering
    \includegraphics[width=\linewidth]{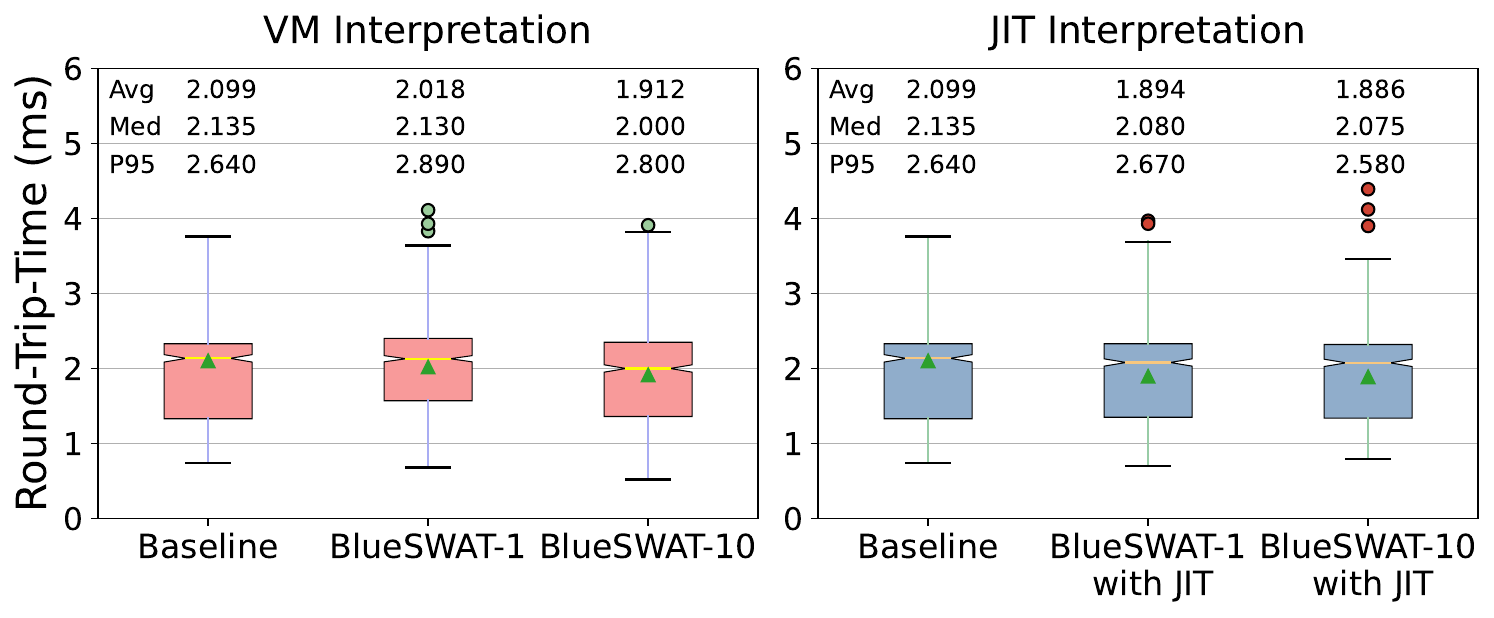}
    \caption{\textbf{RTT (ms) in end-to-end micro benchmark.}}
    \label{fig:microl2ping}
\end{figure}

We further use l2ping~\cite{l2ping} on Linux to perform end-to-end evaluation on \B . We use l2ping to send 1K random L2CAP packets to our nRF52840 DK and evaluate the round-trip time (RTT) of each request. As shown in Figure~\ref{fig:microl2ping}, the average RTT of 1K packets in the baseline is 2.099 ms, which is significantly larger than the latency introduced by \B\ at the microsecond level. In the VM interpretation and JIT mode, the overall distribution of request RTT is similar to the baseline while the average RTT drops. The fluctuation of request RTT almost reaches 4 ms which indicates the latency introduced by different payloads impacts the time consumption more than \B\ on each request. Since not all requests would trigger \B\ inspection and the maximum latency of each rule is less than 5 $\mu s$, the fluctuation at the millisecond level demonstrates that \B\ barely impacts the runtime performance of the stack in end-to-end connection. 
When we add 10 transition rules, the time consumption of 1K requests further drops in both VM interpretation and JIT mode. It is because the first rule we load in \B -1 inspects the link layer packet header field which can be invoked by most packets. The rest of the rules used in \B -10 aim at complicated session-based attacks, such as inspecting redundant SMP keys for the same peer address, which are triggered less frequently than the previous one. Hence, \B\ maintains a stable latency performance when scales to more transition rules against newly disclosed attacks.
In all cases, the interquartile range lies above 1.3 ms and the minimum RTT is higher than 400 $\mu s$. Combined with the above analysis results, the micro benchmark demonstrates that \B\ introduces a negligible latency overhead in end-to-end communications.

\subsubsection{Macro benchmark.}
We test the performance of \B\ in two real-world Bluetooth applications which are Battery Level Service (BAS) and Heart Rate Service (HRS). We use the Nordic Connect for Desktop~\cite{nrfcon} to invoke the services 1K times and record the real-time RTT of the applications. Our test conditions are stricter than real-world scenarios because normally HRS on a smartwatch does not need to update the data frequently. It may take the average heart rate of the user in a 10-second interval while we let the applications continuously refresh and update the values in a 3.5 ms interval. A longer interval in a real-world scenario will relatively reduce the latency impact of \B . 

We load 10 different eBPF programs for both applications and test under VM interpretation and JIT mode. The CDF diagrams of the evaluation results are shown in Figure~\ref{fig:macroben}, Appendix \ref{appendix: benchmark}. For BLS, the average RTT in the baseline is 3489.7 $\mu s$ and P99.5 of RTT is 3740.62 $\mu s$. The baseline time consumption is significantly larger than the latency of \B\ shown in Table~\ref{tab:timetab}, which indicates that \B\ introduces a negligible latency. Under VM interpretation and JIT mode, the average RTT of BAS increases 12 $\mu s$ (0.32\%) which roughly matches the time consumption increase of 10 rules. Running eBPF programs in JIT mode does not explicitly improve the latency performance but requires additional space consumption as shown in Figure~\ref{fig:dram}. Hence, VM interpretation is enough for simple transition rules that we deploy in our prototype.
For HRS, the average RTT in the baseline is 3487.82 $\mu s$ and the P99.5 of RTT is 3736.62 $\mu s$. The time consumption in the baseline is also significantly larger than the latency of \B\ shown in Table~\ref{tab:timetab}, which indicates a minimal impact on HRS. Under VM interpretation and JIT mode, the average RTT of HRS increases around 18.6 $\mu s$ (0.50\%) and 10.8 $\mu s$ (0.29\%). In this case, JIT mode saves around 7.8 $\mu s$ on average application RTT, which roughly matches the time difference in \B -10 and \B -10-JIT shown in Table~\ref{tab:timetab}. In summary, the runtime latency that \B\ introduces to real-world applications is limited to less than 18.6 $\mu s$ (overhead of 0.50\%) and can be considered negligible.

\subsection{Power Performance}

\begin{figure}
\centering
\begin{subfigure}[h]{1\columnwidth}
    \includegraphics[width=1\columnwidth, trim=17 10 0 0,clip]{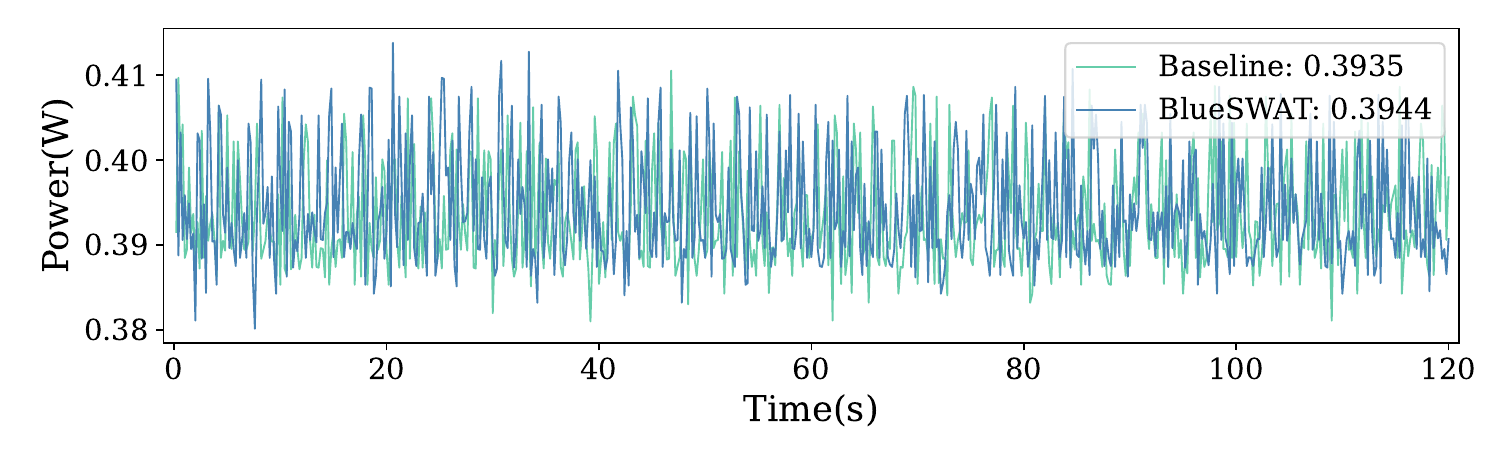}
    \vspace{-15pt}
    \caption{\textbf{Processing power.}}
    \label{fig:power}
\end{subfigure}
\vspace{15pt}

\begin{subfigure}[b]{0.45\columnwidth}
\centering
    \includegraphics[width=\columnwidth, trim=10 0 10 0,clip]{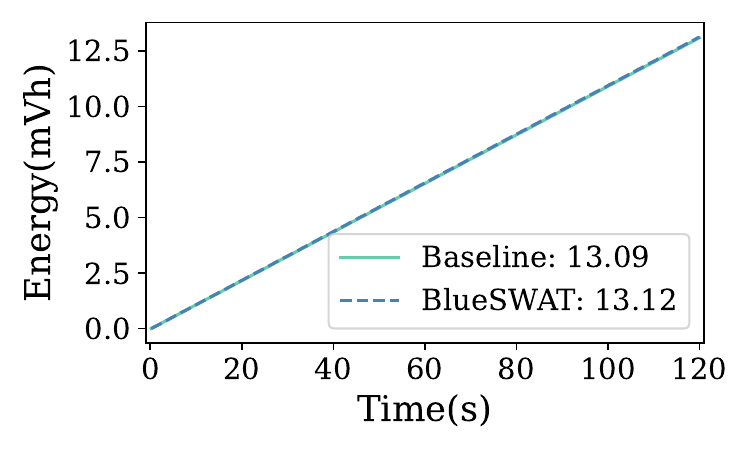}
    \vspace{-17pt}
    \caption{\textbf{Energy usage.}}
    \label{fig:energy}
\end{subfigure}
\hspace{0.05\columnwidth}
\begin{subfigure}[b]{0.45\columnwidth}
\centering
    \includegraphics[width=\columnwidth, trim=10 0 10 0,clip]{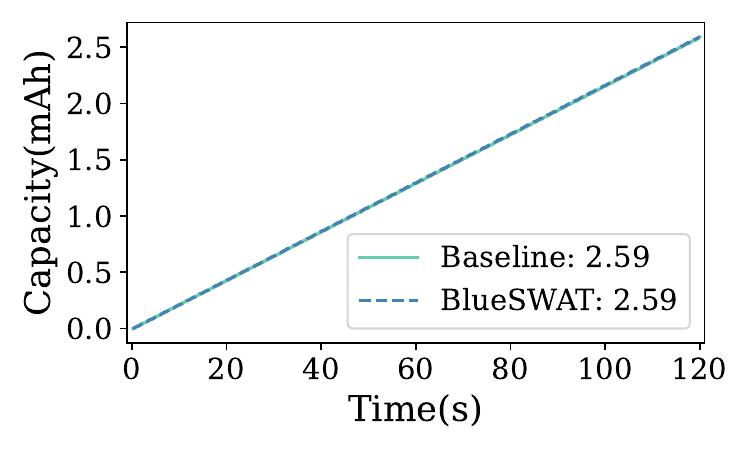}
    \vspace{-17pt}
    \caption{\textbf{Capacity usage.}}
    \label{fig:capacity}
\end{subfigure}

\vspace{10pt}
\caption{\textbf{Evaluation on power performance.}}
\label{fig:power_eva}

\end{figure}

We access the power and energy performance of \B\ over a 120-second window, encompassing four phases: 20s of connection, 40s of BAS, another 20s of connection, and 40s of HRS. To measure real-time processing power, energy use, and capacity of \B , we employed a ChargerLAB POWER-Z KT002 device \cite{kt002} with a sampling rate of 2.5 times per second.  We use a bare Zephyr BLE project as the evaluation baseline. For \B , we load 10 security policies and enable the JIT compiler. The evaluation results are presented in Figure \ref{fig:power_eva}. 

The evaluation reveals that the average power consumption of the baseline is 0.3935 W. \B\ introduces an average of 0.0009 W more power than the baseline, representing a 2.29\% increase. This additional power draw arises from enhanced functionalities of \B . Since the baseline only runs fundamental BLE functions, the power overhead of \B\ will decrease as vendors deploy their real-world applications with a wider range of BLE features. As shown in Figure \ref{fig:power_eva}(b) and \ref{fig:power_eva}(c), the energy and capacity consumption rate of \B\ is almost the same as the baseline. In the 120-second evaluation window, the baseline consumes 13.09 mVh of energy and 2.59 mAh of battery capacity. After deploying \B , the device consumes 13.12 mVh of energy and 2.59 mAh of battery capacity. Nordic 52840 DK uses a Renata CR2032.MFR battery \cite{renata} with a capacity of 260 mAh. Under the circumstances, the battery can support the Nordic board running \B\ for approximately 3.35 hours straight. In a nutshell, \B\ introduces a 0.03 mWh (0.23\%) of energy consumption overhead and minimal overhead on battery capacity usage. Therefore, the impact on device performance and battery life of \B\ is negligible.

\section{Discussion}
\label{sec:discussion}

\textbf{Application on Bluetooth Classic.} \B\ is potentially effective in Bluetooth Classic (BC) defense because BC and BLE share similar protocol architecture and security procedures, such as the link layer (Link Manager in BC) and the pairing procedure. 
Currently, we focus on the BLE protocol because there is no available open-source BC stack on embedded devices for evaluation. Many IoT devices use tailored Linux kernels with BlueZ stack~\cite{bluez}, but they do not openly share their code. 
Research about BC attacks can overcome non-public protocol stack issues because verifying attacking performance does not need firmware source code. But for a defense framework like \B , it requires heavy and tedious efforts to inject vulnerabilities to perform defense evaluation. Nevertheless, we plan to expand \B\ to BC once there are available open-source stacks.

\noindent\textbf{Policy Correctness and Effectiveness.}
As explained in Section \ref{sec:defscale}, a successful mitigation with \B\ demands the \textit{correctness} and \textit{effectiveness} of the policy. Correctness denotes that the policy itself does not include vulnerabilities that will lead to new attacks. Effectiveness signifies that the policy can actually mitigate the targeting attack if it is successfully deployed to the device.

Vendors have the responsibility to make sure that their systems are correct and vulnerability-free. When new vulnerabilities are disclosed, \B\ allows the vendors to rapidly patch it. Therefore, vendors should verify the correctness of patches and ensure the correctness of \B\ policies, which is consistent with existing studies \cite{Tian2019LBMAS, rapidpatch, Chen2018InstaGuardID}. Otherwise, if the policy has fundamentally misinterpreted the attack, the defense is destined to fail. Then, vendors can use our automatic compiler to compile C policies into eBPF bytecode, while \B\ automatically guarantees the correctness of the policies as mentioned in Section \ref{sec:imple}.

A high-level Domain Specific Language (DSL) can potentially facilitate the rule creation process of \B , as shown in Appendix \ref{appendix: dsl}. After vendors identify the transition paths of the attacks, they write C transition rules with the API of \B\ FSM, which provides access to FSM events, states, and variables. The policies of \B\ now contain 20 LOC on average and mostly in a similar format. In Figure \ref{fig:hooks}(b), we present a DSL based on Lua \cite{lua} which can further restrict the policy format and facilitate the development progress. The Lua scripts can be compiled into C with Python first and then compiled into eBPF programs with our compilation tools automatically, which will ensure policy correctness.

\section{Related work}
\label{sec:relatedwork}

\noindent\textbf{BLE Attacks.}
The research on BLE attacks mainly falls into two categories based on the target: firmware compromise \cite{Ruge2020FrankensteinAW, Sivakumaran2018ASO} and Specification exploitation \cite{Tschirschnitz2021MethodCA, Claverie2021BlueMirrorRO, Zhang2020BreakingSP, Antonioli2020BLURtoothEC, Biham2019BreakingTB}. 
SweynTooth \cite{Garbelini2020SweynToothUM} built a general fuzzing framework for BLE implementations and found 24 new vulnerabilities across multiple vendor products. 
BLEEDINGBIT~\cite{bleedingbit} unveiled a set of critical chip-level vulnerabilities affecting BLE chips manufactured by Texas Instruments. It is a typical session-based attack that sends benign packets to store malicious code on the victim device in advance, then overflows the stack with an exploiting packet. 
In recent years, the community also revealed some BLE Specification design flaws that threaten millions of standard-compliant products. 
Antonioli et al. \cite{Antonioli2020KeyND} demonstrated the effectiveness of low key entropy attack on standard compliant BLE devices which is similar to the well-known KNOB \cite{Antonioli2019TheKI} attack on Bluetooth Classic devices. Relay attacks were found capable of circumventing latency bounding or encryption on the BLE link layer and compromising Tesla cars \cite{nccrelay,Xie2023AccessYT}.
BLESA \cite{Wu2020BLESASA} introduced two spoofing attacks targeting the BLE link layer, where an attacker impersonates a previously paired device and feeds forged data to the victim. BLE physical layer was also found vulnerable to signal eavesdropping, injection, and device tracking attacks \cite{linkingble, evaluat, Stute2021DisruptingCO, Stute2019ABO, allowlist, Cayre2021InjectaBLEIM}. Many of the BLE attacks, especially those targeting Specification flaws, are session-based and cannot be detected by existing stateless packet filtering systems.


\noindent\textbf{BLE Security.}
The research on BLE security mainly falls into two categories, which are protocol analysis and defense solutions. 

Formal analysis methods \cite{Wu2022FormalMD, Shi2023FormalAA, Jangid_Zhang_Lin_2023, Troncoso_Hale_2021} built various verification models of the protocol, which mostly focused on the pairing procedure, and uncovered several Specification design flaws. 
However, their proposed mitigation mostly required revising the Specification which resulted in a long patching window, and some were deprecated by Bluetooth SIG due to backward compatibility \cite{btsok, Wu2022FormalMD}. Besides, the traditional software update mechanism demands device reboot and firmware recompilation, which is sometimes impractical for resource-constrained IoT devices. BLEDiff \cite{blediff} proposed a protocol differential testing framework that can identify deviant non-compliant behaviors between multiple BLE implementations. By contrast, we aim to mitigate attacks after vulnerability disclosure via cross-platform instant deployment and stateful traffic inspection.

Existing defense solutions are not as comprehensive as attacking works because a defense framework must consider many restrictions, such as source code availability, backward compatibility, performance overhead, and demand significant engineering workloads. The most comparable solution to \B\ is LBM \cite{Tian2019LBMAS}, a Linux subsystem hardening peripheral protocols (e.g., USB, Bluetooth, NFC) by traffic filtering. Though LBM and \B\ share partly different scopes, in terms of BLE defense, LBM is not capable of monitoring sequential actions of session context, and hence vulnerable to BLE session-based attacks. It also overlooks the link layer which controls the fundamental procedures of BLE connections.
Apart from LBM, 
LightBlue \cite{Wu2021LIGHTBLUEAP} reduced the attack surface of Bluetooth stacks by debloating unneeded code. However, the remaining firmware is still exploitable without comprehensively monitoring the connection messages. 
ProFactory \cite{Wang2022ProFactoryII} can automatically generate secure low-level protocol implementations for IoT protocols in Linux kernels from protocol specifications. However, it only eliminates basic implementation bugs such as memory safety bugs (i.e., buffer overflow, memory leakage, etc.) and concurrency control vulnerabilities (i.e., race and deadlock). The security-critical procedures of BLE, such as pairing and encryption, are left unstudied. BlueShield \cite{Wu2020BlueShieldDS} proposed a defense strategy for spoofing attacks based on physical patterns of BLE packets. Sadly, it needs to deploy additional hardware and must work in a stationary network, which overlooks numerous mobile BLE devices such as wearables.




\noindent\textbf{IoT Peripheral Security.}
Several works \cite{Tian2019LBMAS, Chi2021PFirewallSC, Wang2022ProFactoryII, Li2020T2PairSA} focus on IoT security and take BLE as one of the communication protocols that peripherals may choose, such as USB, WiFi, and Bluetooth. These solutions usually consider the common features of different wireless protocols and hence fail to study BLE security in a fine-grained way. For instance, T2Pair \cite{Li2020T2PairSA} proposed the Universal Operation Sensing which allows IoT users to pair their devices without inertial sensors. However, it ignored the fact that a large amount of BLE attacks happen beyond the pairing phase.
\section{Conclusion}
\label{sec:conclusion}

We propose \B , a lightweight state-aware security framework for IoT devices, which can mitigate BLE session-based attacks. \B\ captures the patterns of session-based attacks and models them as malicious FSM transition paths. During BLE connections, \B\ monitors session sequential actions at runtime, performing an inspection on the session level in addition to separate packets. We develop a lightweight eBPF mechanism to support resource-constrained embedded devices across heterogeneous IoT architectures. For a vulnerability that affects multiple devices, vendors can develop one common patch and distribute it to all victims. Besides, \B\ patches are transmitted as eBPF programs via BLE, which avoids firmware recompilation and device reboot. 
We implement \B\ on 5 real-world devices with different stacks and chips. On our dataset with 101 real-world BLE vulnerabilities, \B\ can mitigate 76.1\% of session-based attacks and 96.4\% of packet-based attacks, which outperforms existing stateless frameworks. In our end-to-end application evaluation, \B\ patches introduce less than 0.08\% of memory overhead on average and negligible latency.

\bibliographystyle{ACM-Reference-Format}
\bibliography{bib}

\appendix
\section*{Appendix}
\appendix

\begin{figure*}[!h]
    \centering
    \includegraphics[width=\linewidth]{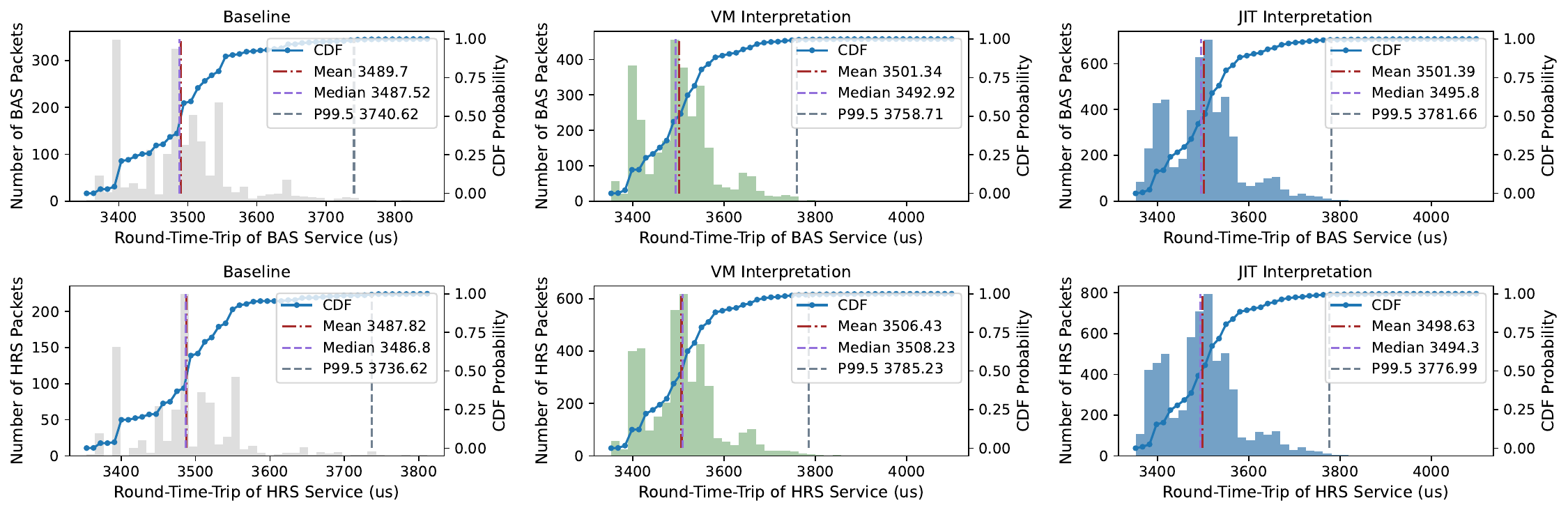}
    \caption{RTT (ms) in real-world applications macro benchmark.}
    \label{fig:macroben}
    \vspace{10pt}
\end{figure*}

\section{Additional Comparison result}
\label{appendix:packet-based}

\begin{table*}[!h]
\vspace{5pt}
    \centering
    \caption{Comparison between \B\ and LBM on mitigating packet-based attacks.} 
    \resizebox{0.6\textwidth}{!}{
    \begin{tabular}{lllcc}
    \toprule[1pt]
        \textbf{Category} & \textbf{CVE} & \textbf{Attack Patterns} & \textbf{LBM} & \textbf{\B} \\
        \specialrule{0.05em}{1pt}{4pt}
        Bounds Check Missing & CVE-2020-10069 & Invalid Channel Map field & \XSolidBrush  &   \Checkmark \\ 
        Buffer Overflow & CVE-2020-10065 & Overflow HCI\_ACL header field & \Checkmark  &  \Checkmark \\ 
        Function Error & CVE-2021-3433 & Invalid Hop Map field & \XSolidBrush & \Checkmark \\ 
        Weak Configuration & CVE-2019-2102 & Hardcoded Long Term Key & \XSolidBrush & \Checkmark \\ 
        Corrupted Pointer & CVE-2022-41972 & Invalid L2CAP channel ID field & \Checkmark  &  \Checkmark \\
        Other (Lack of length check) & CVE-2021-3581 & Overflow SCAN\_REQ payload & \XSolidBrush & \Checkmark \\

     \bottomrule[1pt]
    \end{tabular}
    }
    \label{tab:pktresult}
\end{table*}

\section{Real-world Benchmark Result}
\label{appendix: benchmark}

\section{BLE Connection}
\label{sec:connection}
In this section, we provide some background information about BLE connections.
To establish a secure connection, two devices must go through the following phases. 


\textbf{Discovery.} When the Bluetooth function is enabled, the device first goes into the discovery phase. During discovery, the peripheral (e.g., a headset) starts to send advertising packets, which contain the device address, protocol version, etc. The central (e.g., a smartphone) scans for advertising packets in radio range, and sends out connection requests to target devices. In the discovery phase, the two devices exchange several supporting features and establish a link layer connection at last.

\begin{figure}[h]
    \centering
    \includegraphics[width=1\linewidth,trim=10 0 15 10,clip]{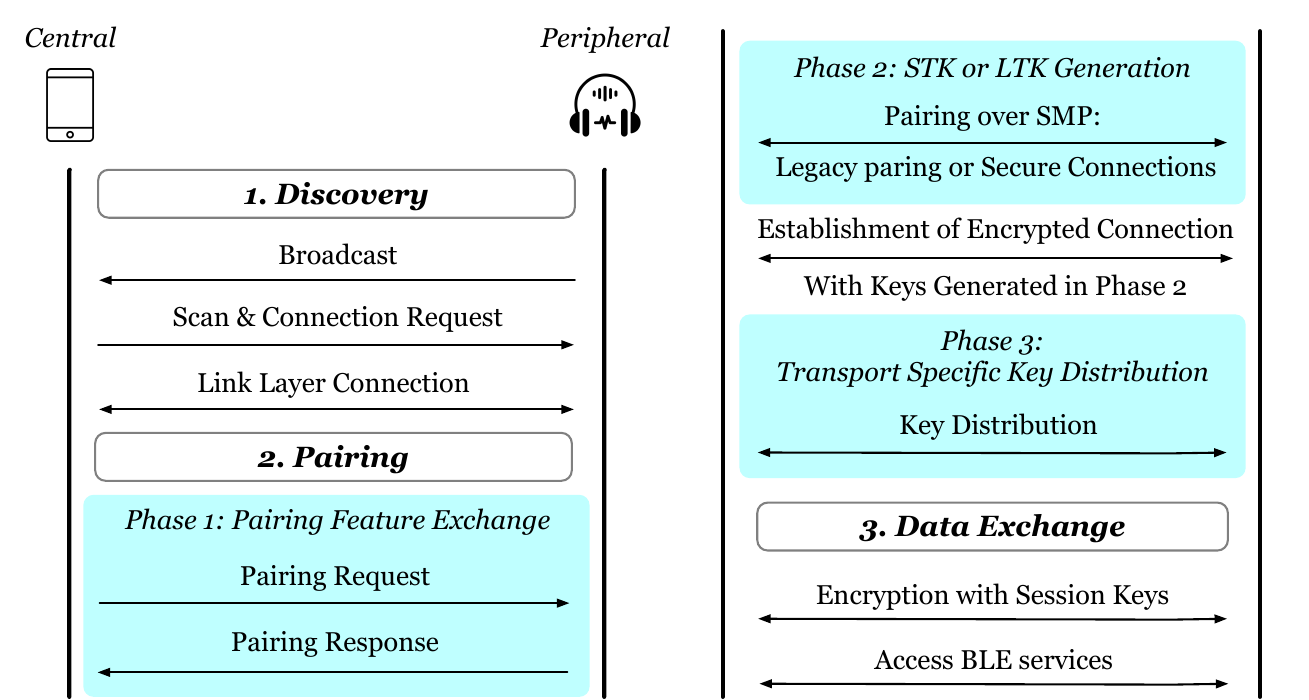}
    \caption{BLE connection process.}
    \label{fig:back}
\end{figure}

\textbf{Pairing.} After establishing a LL connection, the central and peripheral try to negotiate and distribute a shared secret key with the SMP pairing procedure. The pairing procedure includes three phases, which are Pairing Feature Exchange (Phase 1), Short Term Key (STK) or Long Term Key (LTK) Generation (Phase 2), and Transport Specific Key Distribution (Phase 3).

In Phase 1, the devices first exchange authentication requirements and IO capabilities to determine the association models and pairing methods used in Phase 2. OOB authentication data availability, key size requirements, and which transport specific keys to distribute are also exchanged in Phase 1.

In Phase 2, the Bluetooth Specification provides two pairing protocols, which are the \textit{Secure Connections Pairing} (SCP) and \textit{legacy pairing}. Released in Bluetooth v4.0, the legacy pairing protocol generates two keys, which are Temporary Key and STK, to encrypt the connection. Since it does not use the Diffie-Hellman Key Exchange (DHKE) protocol during key sharing, legacy pairing is vulnerable to passive attacks. In Bluetooth v4.2, SCP is introduced to mitigate the weakness by adopting the DHKE protocol. 
There are four pairing methods to use during pairing: \textit{Just Works}, \textit{Out of Band}, \textit{Passkey Entry}, and \textit{Numeric Comparison}. The methods with higher security levels involve user interaction to enhance the authentication procedure, which can mitigate eavesdropping and the MITM attack. SCP is found vulnerable to reused passkey attacks and reflection attacks and receives mitigations through Specification updates. At the end of Phase 2, two devices establish an encrypted connection with the shared STK or LTK.

In phase 3, two devices may distribute transport specific keys, such as the Identity Resolving Key (IRK) value and Identity Address information. Phase 1 and Phase 3 are identical regardless of the method used in Phase 2. 

\textbf{Data Exchange.} Once the central and peripheral exchange the sharing keys, devices can start the encryption and authentication procedure. To derive session keys from Link Keys, BLE performs a proactive or reactive authentication procedure. The session keys are then used for connection encryption and are refreshed every time the devices reconnect.

\section{Domain Specific Language}
\label{appendix: dsl}

\begin{figure}[!h]
\vspace{-3pt}
    \centering
    \begin{subfigure}[b]{1\columnwidth}
        \begin{lstlisting}[language=C]
#define BT_KEYS_LTK_P256 32
#define BT_KEYS_LTK 4
#define BT_KEYS_AUTHENTICATED 1

uint64_t zephyr_filter(uint8_t *newState)
{
	struct FsmState *fsm = (struct FsmState *)newState;

	// Check that if a new pairing procedure with an existing bond will not lower the established security level of the bond.

	if (!(fsm->dc_param[SMP_KEYS] & (BT_KEYS_LTK_P256 | BT_KEYS_LTK))) {
		return IFW_OPERATION_PASS;
	}

	if (fsm->dc_param[SMP_ENC_SIZE] > fsm->dc_param[SMP_ENC_SIZE_PREV]) {
		return IFW_OPERATION_REJECT;
	}

	if ((fsm->dc_param[SMP_KEYS_FLAGS] & BT_KEYS_AUTHENTICATED) &&
	    fsm->dc_param[SMP_METHOD_PREV] == JUST_WORKS) {
		return IFW_OPERATION_REJECT;
	}

	return IFW_OPERATION_PASS;
}
        \end{lstlisting}
        \caption{An example of \B\ C policy.}
        \label{code:BlueSWAT C policy.}
    \end{subfigure}
    
    \vspace{15pt}
    \begin{subfigure}[b]{1\columnwidth}
    \begin{lstlisting}
rule1 =  (bit.band(DC[SMP_KEYS] or 0, KEYS_LTK_P256 | KEYS_LTK) == 0) 
rule2 = (DC[SMP_ENC_SIZE] <= DC[SMP_ENC_SIZE_PREV])
rule3_not_just_work =  (bit.band(DC[SMP_KEYS_FLAGS] or 0, KEYS_AUTHENTICATED) == 0 or
     DC[SMP_METHOD_PREV] ~= "JUST_WORKS")

return rule1 and rule2 and rule3_not_just_work
        \end{lstlisting}

        \caption{Policy in Lua. Lines 1-3 are customized user variables and Lines 9-14 are user-defined rules.}
        \label{code:dsl}
    \end{subfigure}
    
    \vspace{10pt}
    \caption{\B\ C policy and Lua policy.}
    \label{fig:hooks}
    \vspace{-2pt}
\end{figure}

\section{BLE vulnerabilities}
\label{appendix:vulsta}

\begin{figure}[!h]
\vspace{-10pt}
    \centering
    \includegraphics[width=1\linewidth, trim=10 10 0 0,clip]{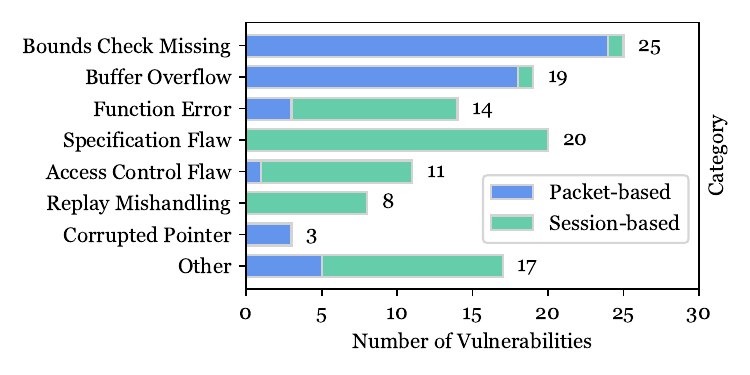}
    \caption{Category of BLE Vulnerabilities. Vulnerabilities under the "Other" category are strongly related to specific coding errors.}
    \label{fig:vul-stat}
\end{figure}

In this section, we provide the investigation result of BLE vulnerabilities. We systematically collect 117 BLE vulnerabilities in our CVE dataset by November 2023 (as shown in Figure \ref{fig:vul-stat}, Appendix \ref{appendix:vulsta}). Specifically, we investigate a wide range of vulnerabilities which include 94 CVEs marked with "ble" and "bluetooth low energy" from the CVE database \cite{CVE-base}, 16 documented CVEs of open-source BLE stacks \cite{zephyrbt, nimble} and the vulnerabilities disclosed by previous publications \cite{Sivakumaran2018ASO, Xie2023AccessYT, Wu2020BLESASA, Claverie2021BlueMirrorRO, Zhang2020BreakingSP, linkingble, Antonioli2020KeyND, Shi2023FormalAA, Tschirschnitz2021MethodCA, Garbelini2020SweynToothUM, allowlist}.
We aim to protect the firmware layer and host layer of the BLE protocol \cite{btsok}. Attacks \cite{Koh2022BLAPBL, allowlist, linkingble, Cayre2021InjectaBLEIM, Xie2023AccessYT} on the BLE physical layer and application layer, such as signal injection and device tracking, are out of the scope. 

Around 54\% of the vulnerabilities are session-based, which is left unstudied by previous research. Specifically, all Specification weaknesses are session-based because they need complicated interactions to be exploited. These weaknesses cause severe consequences, such as MITM session hijacks and encryption breaks, which threaten all BLE devices. Besides, Function Errors and Access Control Flaws that are caused by implementation faults are usually session-based. To exploit these vulnerabilities, the attackers need to carefully craft malicious message sequences. Replay attacks are obviously session-based because they manipulate duplicate messages. Stateless packet filter mechanisms like LBM cannot detect these attacks since the individual message is usually benign.

\end{document}